\documentclass[structabstract]{aa}

\usepackage{graphicx}
\usepackage{txfonts}
\usepackage{natbib}

\def\lya{\mbox{Ly$\alpha$}}
\def\wb{\mbox{$B_{435}$}}
\def\wv{\mbox{$V_{606}$}}
\def\wi{\mbox{$i_{775}$}}
\def\wz{\mbox{$z_{850}$}}
\def\lesssim{\mathrel{\hbox{\rlap{\hbox{\lower4pt\hbox{$\sim$}}}\hbox{$<$}}}}
\def\gtrsim{\mathrel{\hbox{\rlap{\hbox{\lower4pt\hbox{$\sim$}}}\hbox{$>$}}}}

\begin{document}

\title{The unusual N\,{\sc iv}]-emitter galaxy GDS~J033218.92-275302.7:\\
star formation or AGN-driven winds from a massive galaxy at z=5.56}

\author{E. Vanzella\inst{1}
     \and
      A. Grazian\inst{2}
      \and
      M. Hayes\inst{3}
      \and
      L. Pentericci\inst{2}
      \and
      D. Schaerer \inst{3,4}
      \and
      M. Dickinson\inst{6}
      \and
      S. Cristiani\inst{1}
      \and
      M. Giavalisco\inst{7}
      \and
      A. Verhamme\inst{5}
      \and
      M. Nonino\inst{1}
      \and
      P. Rosati\inst{8}
      }

\institute{
       INAF - Osservatorio Astronomico di Trieste, Via G.B. Tiepolo 11,
       40131 Trieste, Italy.
       \and
       INAF - Osservatorio Astronomico di Roma, Via Frascati 33,
         I-00040 Monteporzio Roma, Italy
       \and
       Geneva Observatory, University of Geneva, 51, Ch. des Maillettes, CH-1290 Versoix, Switzerland
       \and      
       Laboratorie d'Astrophysique de Toulouse-Tarbes, Universit\`e de Toulouse, CNRS, 14 Avenue E. Belin, F-31400 Toulouse, France
       \and
       Department of Physics, University of Oxford. Denys Wilkinson Building, Keble Road, Oxford, UK
       \and
       NOAO, PO Box 26732, Tucson, AZ 85726, USA
       \and
       Astronomy Department, University of Massachusetts, Amherst MA
  01003, USA
       \and
       European Southern Observatory, Karl-Schwarzschild-Strasse 2,
       Garching, D-85748, Germany.
     \thanks{Based on observations made at the European Southern
Observatory, Paranal, Chile (ESO program 170.A-0788 {\it The Great
Observatories Origins Deep Survey: ESO Public Observations of the
SIRTF Legacy/HST Treasury/Chandra Deep Field South.}) }
       }

\offprints{E. Vanzella, \email{vanzella@oats.inaf.it}}

\date{Received - / Accepted -}

\abstract{}{We investigate the nature of the source GDS~J033218.92-275302.7 
at redshift $\sim$ 5.56.} 
{The spectral energy distribution of the source is well-sampled by 16 bands 
photometry from UV-optical (HST and VLT),
near infrared, near infrared (VLT) to mid-infrared (Spitzer).
The detection of a signal in the mid-infrared Spitzer/IRAC bands 5.8, 8.0 $\mu m$ -- where 
the nebular emission contribution 
is less effective -- suggests that there is a Balmer break, the signature of
an underlying stellar population formed at earlier epochs.
The high-quality VLT/FORS2 spectrum shows a clear \lya\ emission line,
together with semi-forbidden N\,{\sc iv}] 1483.3-1486.5 also in emission.
These lines imply a young stellar population.
In particular, the N\,{\sc iv}] 1483.3-1486.5 feature (if the source is not hosting an AGN)
is a signature of massive and hot stars with an associated nebular emission.
Conversely, it may be a signature of an AGN.
The observed SED and the \lya\ emission line profile 
were modeled carefully to investigate the internal properties of the source.}
{From the SED-fitting with a single and a double stellar population and from the \lya\
modeling, it turns out that the source seems to have an evolved
component with a stellar mass of $\sim$5$\times$10$^{10}$ $M_{\odot}$ and age
$\sim$ 0.4 Gyrs, and a young component with an age of $\sim$ 0.01 Gyrs and 
star formation rate in the range of 30-200 $M_{\odot}yr^{-1}$.
The limits on the effective radius derived from the ACS/z850 and VLT/Ks bands 
indicate that this galaxy is denser than the local ones with similar mass.
A relatively high nebular gas column density is favored from the \lya\ line modeling
($N_{HI}$$\gtrsim$$10^{21}$$cm^{-2}$). 
A vigorous outflow ($\sim$ 450 km/s) has been measured from the optical spectrum,
consistent with the \lya\ modeling. 
From ACS observations it turns out that the region emitting \lya\ photons is
spatially compact and has a similar effective radius ($\sim$ 0.1 kpc physical)
estimated at the $\sim$1400\AA~rest-frame wavelength, whose emission is dominated 
by the stellar continuum and/or AGN.
The gas is blown out from the central region, but, given the mass of the
galaxy, it is uncertain whether it will pollute the IGM to large distances. 

We argue that a burst of star formation in a dense gas environment is active
(possibly containing hot and massive stars and/or a low luminosity AGN), 
superimposed on an already formed fraction of stellar mass.}{}

\authorrunning{Vanzella et al.}
\titlerunning{The unusual N\,{\sc iv}]-emitter galaxy at z=5.56.}
\maketitle

\section{Introduction}

In the past few years, dedicated space-borne and ground-based observatories and 
refined techniques have allowed us to discover and analyze galaxies 
at increasingly large distances. It is common practice
in observational cosmology to select efficiently star-forming galaxies 
(e.g. Lyman break galaxies, LBGs,  or Lyman alpha emitters, LAEs) and active galactic
nuclei (AGN) up to redshift 6.5 \citep[e.g.][]{Ste99,dick04,Giava04a,
bouwens06,tanigu05,ando06,crist04,fan03,fontanot07}
or spheroidal/fossil massive galaxies up to redshift 4, whose apparent morphology has 
recently introduced a sub class of $ultradense$ objects 
\citep[e.g.][]{daddi05,cimatti08,dokkum08,bui08}.

High-redshift galaxies have been studied through deep 
multi-wavelength surveys, with the aim of maximizing the information on the
energetic output of the sources spanning a wide range of the electromagnetic spectrum 
from X-ray to radio wavelengths. 
This approach has shown its efficiency of constraining luminosity functions 
up to redshift 6-7 and down to few percent of $L^{*}$ (e.g. \cite{bouwens07,bouwens08}),  
stellar mass functions (for masses higher than $10^{9}M_{\odot}$) up 
to redshift 6 \citep{fontana06,fontana09,eyles05,eyles07,stark07,stark09,yan05,labbe06}
and morphology evolution \citep{tanigu09,conse08,ferg04}.

Detailed studies of internal the properties of high-redshift 
galaxies, such as information about hot stars, dust, 
ionized gas in $H II$ regions, and the large-scale outflows of neutral and 
ionized interstellar material (ISM) are now becoming feasible up to redshift 4-5 
\citep[e.g.][]{shapley03,ando07,ouchi08,vanz09}. 
Useful information from the \lya\ profile modeling of high-redshift galaxies with 
radiative transfer codes is producing interesting constraints on the dynamical and 
physical state of the ISM and ionizing sources (e.g. \cite{ver08}; \cite{sch08}).

From this point of view, thanks to the combination of depth, area, and multivawelength coverage,
the Great Observatories Origins Deep Survey project 
(see \cite{dick03b}; \cite{giava04b} for a review about this project)
is ideal for studing galaxies at high redshift and the connection between photometric, 
spectroscopic, and morphological - size properties
\citep[e.g.][]{pente07,pente09,conse08,ravi06,vanz09}, and 
their relation with the environment \citep[e.g.][]{elbaz07}.

Along with enabling a systematic study of normal galaxies, multi-frequency surveys
over large areas and depth also allow us to discover rare objects.
In particular, a new class of objects showing prominent 
N\,{\sc iv}] 1486 emission have recently been reported. Such a feature is rarely seen at any redshift. 
A small fraction (1.1\%, 1.7$<$z$<$4)
of the QSO sample extracted from the SDSS fifth data release is nitrogen rich, showing
N\,{\sc iv}] 1486 or N\,{\sc iii}] 1750 emission lines and N\,{\sc v}] 1240 also in emission 
(typically stronger than 
the rest of the population \citep{jiang08}.
Similarly, \cite{glikman07} discuss the discovery of two low-luminosity quasars at redshift
$\sim$ 4 with \lya\ and C\,{\sc iv} lines, moderately 
broad N\,{\sc iv}] 1486 emission, and an absent N\,{\sc v}] 1240 line.
In these particular cases, the blinding intensity of the central engine is reduced,
allowing study of the properties of the host galaxy.
\cite{fosbu03} report on an $HII$ lensed galaxy at redshift 3.357 (the $Lynx~arc$) 
whose spectrum shows N\,{\sc iv}] 1486, O\,{\sc iii}] 1661, 1666, C\,{\sc iii}] 1907, 1909, as well as 
the absence of the N\,{\sc v}] 1240 line. Their modeling of the spectrum favors
a hot (T$\sim$80000K) blackbody over an AGN as  the ionizing source. Alternatively,
\cite{binette03} suggest an obsured AGN as a photoionizing source of the Lynx arc.
\cite{VM04} propose a population of Wolf-Rayet (WR) stars as the ionization source for the same 
$HII$ galaxy,
with an age below 5 Myr that contributes to a fast enrichment of the interstellar medium.
In this scenario the stars involved are much colder than those proposed in \cite{fosbu03}.
 
In the present work we report on the source GDS J033218.92-275302.7 at redshift 5.563
located within the GOODS southern field, for which extensive information 
(photometry, spectroscopy and morphology) is available. The galaxy has been discovered
during the ESO/FORS2 spectroscopic survey (\cite{vanz06}).
We focus our attention on this source because it shows several unique characteristics.
First the high S/N spectrum exibits a relatively bright N\,{\sc iv}] 1486 feature 
in emission, a unique example among the more than 100 spectra of high z starburst 
galaxies that were obtained from the GOODS/FORS2 campaign (\cite{vanz08}).
Second, while the spectrum shows a bright \lya\ line indicating
a young stellar component, the photometry shows a prominent 
Balmer break indicating that there is also an evolved component.
Finally, the bright IRAC flux suggests that this is a massive
galaxy, especially interesting if one considers its very high redshift.

Indeed the object was already noted by several authors; for example, 
\cite{fontanot07} selected it as an AGN candidate on the basis of 
morphological and color considerations, but discarded it from the sample
because of its peculiar $galaxy-like$ optical spectrum.
\cite{wik08} report it among the 11 candidates with 
photometric redshifts in the range 4.9$<$z$<$6.5, dominated by an old 
stellar population, with ages 0.2-1.0 Gyr and having very high 
stellar masses, in the range $(0.5-5)\times 10^{11}$ $M_{\odot}$.
Also \cite{stark07} report for this source a stellar mass
highr than $10^{11}~M_{\odot}$. Similarly  \cite{pente09} 
note that this is one of a handful of bright \lya\ emitting LBGs 
at high redshift with an evolved population, indicating that not all Lyman alpha
emitters are young primeval galaxies.

As noted recently by \cite{sch09} and \cite{raiter09}, strong nebular emission lines 
may bias the result of the SED-fitting of high-redshift galaxies 
(e.g. [O\,{\sc iii}] 4959-5007, H$\alpha$). Indeed the apparent photometric breaks might actually be 
produced in some cases by the boost of some of the lines.
Something similar may be happening in this case, therefore
it is important to quantify the strength of these lines and their influence on global photometry.

In the present work we perform dedicated SED-fitting allowing
single and multiple stellar populations, and important information is extracted from 
the \lya\ profile modeling. Together with the morphological appearance, constraints 
have been placed on the stellar mass density, ages, gas, dust content, and outflows.

The work is structured as follow. In  Section 2 a summary of the 
photometric, spectroscopic, and morphological properties is given, and in Sect. 3
the possible scenarios are discussed about the nature of the source. Section 4 
describes the SED and \lya\ modeling, and in Sect. 5 we discuss the results.
Section 6 concludes the work. In the following the standard cosmology is
adopted ($H_{0}$=70 km/s/Mpc, $\Omega_{M}$=0.3,$\Omega_{\Lambda}$=0.7). 
If not specified, magnitudes are given on the AB system.

\section{Source GDS~J033218.92-275302.7}

Source GDS~J033218.92-275302.7 is located within the southern field 
of the Great Observatories Origins Deep Survey. 
The multi-wavelength observations consists of deep 
U, R (VLT), \wb\ , \wv\ , \wi\ and \wz\ (HST), $Js$, $H$, $Ks$
(VLT), 3.6, 4.5, 5.8, 8.0, and 24 $\mu m$ (Spitzer) bands. Moreover observations
in the X-ray and radio domains are available from Chandra and the Very Large 
Array, respectively \citep{luo08,miller08}.

A considerable part of the spectroscopic information of that field has been collected 
by the VLT/FORS2 spectrograph, which has produced about one thousand 
redshift determinations (with a resolution of $\sim$13\AA, at $\sim$8600\AA), 
between redshift 0.5 and 6.2, in particular more than one hundred LBGs have 
been confirmed at redshift beyond 3.5 \citep{vanz05,vanz06,vanz08}. 
The VLT/FORS2  spectroscopic survey has been complemented in the redshift 
interval 1.6 $<z<$ 3.5 and at $z<$ 1 by the VLT/VIMOS spectroscopic survey, which is 
producing more than 5000 spectroscopic identifications 
\citep{popesso08,balestra10}. Such measurements increase the
spectroscopic information available from previous works \citep[e.g. VVDS][]{fevre05,szo04}.

Source GDS~J033218.92-275302.7 with \wz=24.61$\pm$0.03
was selected as a \wv--band dropout and was confirmed to be 
at redshift 5.563 (redshift of the \lya\ line, \cite{vanz06}). 

\subsection{UV spectral properties}
\begin{figure}[] \centering
  \resizebox{\hsize}{!}{\includegraphics{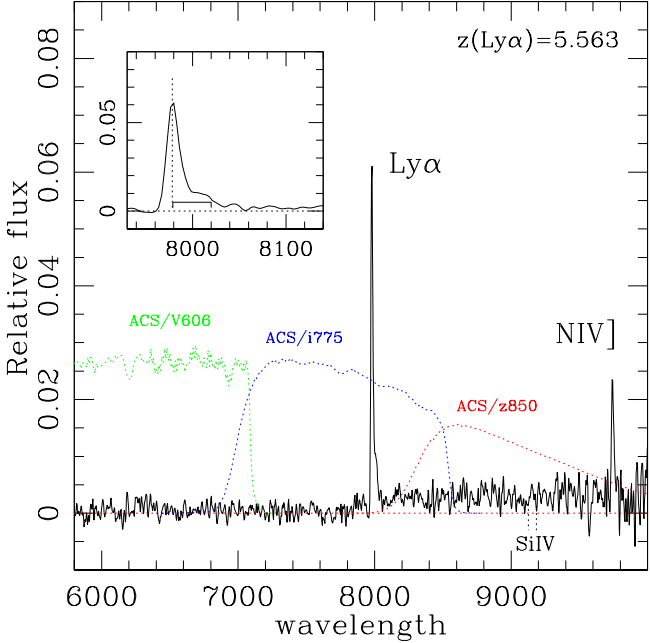}}
\caption{1-dimensional spectrum of the galaxy discussed in the present work.
The \lya\ and N\,{\sc iv}] 1486 emission lines are evident, together with the detection
of the continuum and the IGM attenuation. 
The inner box shows a zoom of the \lya\ line, the asymmetry and the red tail are 
visible (the latter is marked with a solid segment, 40\AA~long, or 1500 $km~s^{-1}$).
The transmissions of the ACS \wv\ , \wi\ , and \wz\ are also shown. The exposure time was 14400s.}
\label{fig3}
\end{figure}

The main spectral features are the \lya\ emission line (EW$\sim$60\AA~rest frame),
the break of the continuum blueward of the line, and the semi-forbidden emission line 
N\,{\sc iv}] 1486, a doublet $\lambda\lambda$
1483.3-1486.5\AA~(see Fig.~\ref{fig3} and Table~\ref{tab1} for a physical 
quantities summary).
The detection of the N\,{\sc iv}] 1486 in emission is unusual for LBGs. However, this
atomic transition has been identified by \cite{fosbu03} in the $Lynx$ arc
and in a sub class of QSOs (e.g., \cite{glikman07}, \cite{baldwin03}).
In the following the main properties of the UV spectrum are described:
\begin{table}
\centering 
\caption{Summary of the physical quantities derived from the spectral features
and morphological analysis.} 
\label{tab1} 
\begin{tabular}{l l} 
\hline\hline 
Obs. Spect. properties &\\ 
\hline 
z ( \lya\ 1215.7)     & 5.563 \\ 
FWHM ( \lya\ 1215.7) & 600$\pm$100 $km~s^{-1}$ \\
$EW_{0}$( \lya\ )   & 59$^{+195}_{-29}$\AA~(89\AA~from phot.$^{\star}$)\\
L( \lya\ )         & 3.8$^{+0.3}_{-0.3}$$\times$10$^{43}$$erg~sec^{-1}$\\
$SFR_{0}$( \lya\ )  & 31 $M_{\odot}/yr$\\
\hline
z (N\,{\sc iv}] 1486)      & 5.553 (1486.5\AA)\\
FWHM (N\,{\sc iv}] 1486) & 400$\pm$100 $km~s^{-1}$ \\
$EW_{0}$(N\,{\sc iv}] 1486)& $22^{+64}_{-10}$\AA~(33\AA~from phot.$^{\star}$)\\
L(N\,{\sc iv}] 1486) & $1.3^{+0.3}_{-0.4}$$\times$10$^{43}$$erg~sec^{-1}$)\\
\hline
L(N\,{\sc v} 1240-1243)   & $<$1.06$\pm$0.57 $\times$10$^{42}$$erg~sec^{-1}$ \\ 
\hline
L(Si\,{\sc iv} 1393.8-1402.8) & $<$1.13$\pm$0.73$\times$10$^{42}$$erg~sec^{-1}$ \\ 
\hline
L(C\,{\sc iv}] 1548.2-1550.8) & $<$6$\times$10$^{42}$$erg~sec^{-1}$ (2$\sigma$ limit)$\dag$\\
\hline
V(\lya\ - N\,{\sc iv}] 1486) & +457$\pm$50$km~s^{-1}$ \\ 
\hline
\hline
Morph. (rest-frame)  & (kpc physical)\\    
\hline
$r_{e}~(\lya)$ (GALFIT) & 0.08$\pm$0.01 kpc (ACS \wi\ band) \\
$r_{e}~(1400\AA)$ (GALFIT) & 0.11$\pm$0.01 kpc (ACS \wz\ band) \\
$r_{e}~(3300\AA)$ & $<$ 0.9 kpc (0.6'' seeing, ISAAC $Ks$ band)\\ 
$S/G~(1400\AA)$ (SExtr.)& 0.83 (S/G, ACS \wz\ band)\\
$area~(1400\AA)$ (SExtr.)& 8.8 kpc$^{2}$ (AREAF 303 pix., \wz\ band)\\
\hline 
\multicolumn{2}{l}
{$\dag$ continuum not detected, 2$\sigma$ of the noise fluctuations is adopted.}\\
\multicolumn{2}{l}
{$\star$ adopting the continuum derived from the ACS \wz\ band.}\\
\end{tabular}
\end{table}
\begin{figure}[] \centering
  \resizebox{\hsize}{!}{\includegraphics{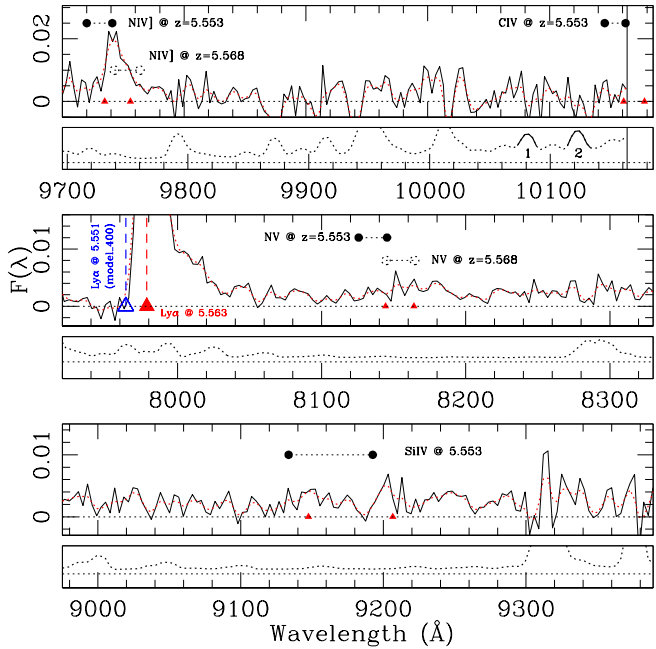}}
\caption{Three zoomed regions of the 1D spectrum (red dotted line is the solid black line 
smoothed over two pixels). Dotted plots show the rescaled sky
spectrum. Filled circles mark the position of the lines at the redshift 
5.553 (case in which only the N\,{\sc iv}] 1486.5 component is detected). Dotted open circles 
denote the other (but less probable) case in which the N\,{\sc iv}] 1483.3 component is detected
(z=5.568).
Red filled triangles mark the positions at the observed \lya\ redshift, z=5.563.
{\bf Top:} the positions of the N\,{\sc iv}] 1486 and C\,{\sc iv}] features are shown. In the sky spectrum two
sky emission lines are marked: (1) 10082.46\AA~and (2) 10124.01\AA~(also shown in the 2-dimensional spectrum
of Fig.~\ref{fig_end_spec}). 
{\bf Middle:} the region where N\,{\sc v} is expected to lie is shown. The open blue triangle indicates the 
estimated $intrinsic$ \lya\ redshift from the \lya\ modeling.
{\bf Bottom:} The region around the SiIV feature is shown.
In the $middle$ ($\lambda$$\sim$8150\AA) 
and $bottom$ ($\lambda$$\sim$9170\AA) cases, the sky-lines contamination is minimal.}
\label{fig_zoom}
\end{figure}
\begin{enumerate}
\item{As shown in the top panel of Fig.~\ref{fig_zoom} an emission line at $\lambda$ 9742\AA~is 
detected. 
The spectral resolution is in principle sufficient to resolve the double profile 
of the two N\,{\sc iv}] 1486 components, i.e 1483.3\AA~ and 
1486.5\AA. 
We interpret this feature as the detection of one of the two components. 
A first possibility is that this line is the 1483.3\AA~component: 
in this case the redshift turns out to be higher than the observed \lya\ redshift. 
The \lya\ emission from LBGs is commonly observed to be redshifted relative to the
systemic velocity traced by other, non-resonant emission lines or stellar absorption lines
(\cite{shapley03}, \cite{tapken07}, \cite{ver08}, \cite{vanz09}). Therefore, it would be 
unusual if, in this object, the N\,{\sc iv}] redshift were higher than that from \lya\ .
The other possibility is that we are detecting the 1486.5\AA~component 
at redshift 5.553, suggesting a high-density 
limit\footnote{the ratio of those components is related to the electron 
density, \cite{raiter09}, \cite{keenan95}}
and a velocity offset (i.e. a presence of an outflow) 
between \lya\ and N\,{\sc iv}] lines of +457 km/s (dz=0.01).
A further proof of this possibility is that the redshift and 
outflow estimated from the spectrum are consistent with the results 
from the \lya\ profile modeling discussed below (see Sect. 4.2). Therefore, 
in the following we assume the line to be the 1486.5\AA~component.}
\item{There is no detection for the SiIV 1393.8-1402.8 doublet, 
either in emission or absorption, and similarly for the N\,{\sc v} 1240-1243 doublet;
their luminosity limits are reported in Table~\ref{tab1} (see also Figure~\ref{fig_zoom}).}
\item{From the 2-dimensional spectrum, the FWHM of the spatial profiles
of the \lya\ and the continuum (by collapsing columns along the Y-axis, see Fig.~\ref{fig2})
are fully comparable, $\sim$ 0.7 arcsec. This is also compatible
with the seeing during observations, 0.7 arcsec. More interestingly,
a better constraint on the \lya\ extension comes from the ACS \wi\ band.
As shown in Fig.~\ref{fig3}, the \wi\ band is mainly probing the UV emission
region between the \lya\ line and 1300\AA~(less than 100\AA~rest-frame), and the part 
blueward of the \lya\ line is strongly attenuated by the IGM absorption.
Since the \lya\ equivalent width is $\sim$ 60\AA, it turns out that the
ACS \wi\ image is dominated by the \lya\ emission. In Sect. 2.3 we show that the
morphological properties derived from the \wi\ band (probing the \lya\ line) and the \wz\ band 
(not containing the \lya\ line and probing the emission at 1400\AA) are similar, 
e.g. the effective radii of the two are the same order. This implies that the spatial extension of 
the \lya\ line is similar to the emitting region at 1400\AA~rest-frame.}
\end{enumerate}

Figure~\ref{fig2} shows the contour plots of the 2-dimensional \lya\ region.
Various sigmas above the noise fluctuation are reported from 1 to 20. It is 
evident that the asymmetric shape in the wavelength domain (that extends at 
least $\sim$ 40\AA~(3$\sigma$)). If the \lya\ emission arises from a simple expanding
shell of material, then it is expected to have less of velocity width
at its outer extremes than along the line of sight to the central region,
where the transverse velocity component tends to zero.

\begin{figure}[] \centering
  \resizebox{\hsize}{!}{\includegraphics{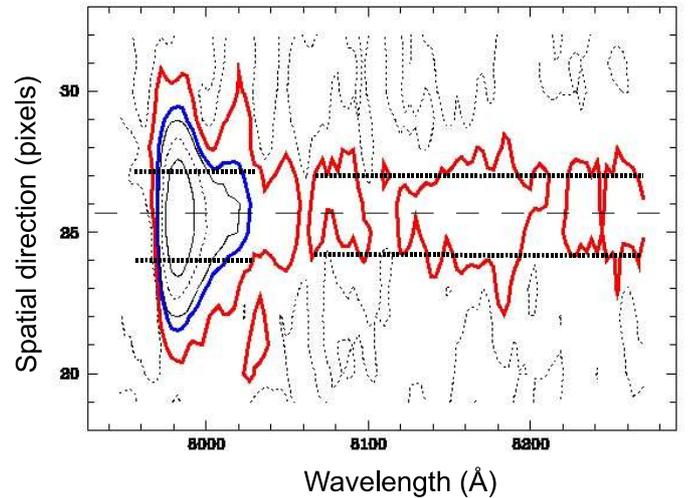}} 
\caption{Contour plot of the 2-dimensional \lya\ region, spatial
direction (pixels, 1 pixel=1.5 kpc) versus wavelength (\AA~, 20\AA~corresponds to 751 $km~s^{-1}$ ). 
The spectral interval of 7960-8280\AA~and spatial extension of 22.5 kpcs are shown. 
Thin dotted and continuous black lines mark
0$\times$, 5$\times$, 10$\times$, 20$\times$$\sigma$ above the mean signal. Thick
lines mark 1$\sigma$ (red) and 3$\sigma$ (blue). Horizontal dotted lines indicate 
the FWHM of the profiles (spatial extension) by collapsing columns in the \lya\ region and on the
continuum region.} 
\label{fig2}
\end{figure}
\subsection{Photometric properties}

Fig.~\ref{fig_SED} shows the overall SED (black squares) and
Table~\ref{tab3} summarizes the multi-band photometry of the source
collected from different instruments mounted on ground and space-based telescopes.
Magnitudes, errors and 1-$\sigma$ lower limits (l.m.) are derived from the MUSIC catalog, 
\citep{grazian06,santini09}.  There are other two WFI U bands observations not reported in the table, 
U35 and U38, with a slightly different filter shape. Their lower limits are 27.84 and 26.75, 
respectively, much shallower than the limit provided by the VLT $U_{VIMOS}$.

Despite the high redshift of the source discussed here, its (\wi-\wz) color 
is 0.59, significantly bluer than the typical threshold of $\sim$1.3 adopted 
to select galaxies beyond redshift $\sim$ 5.5 \citep[e.g.][]{dick04,bouwens07}. 
This comes from the contribution of the \lya\ emission to the flux in the \wi\
filter (see Fig.~\ref{fig3}), that decreases
the (\wi-\wz) color by about 0.7 magnitudes (as shown in \cite{vanz09}).
For this reason it has been selected as a \wv--band dropout source.
It is also an R--band dropout source if referred to the ground-based 
photometry, see Table~\ref{tab3}.
\begin{figure}[] \centering
  \resizebox{\hsize}{!}{\includegraphics{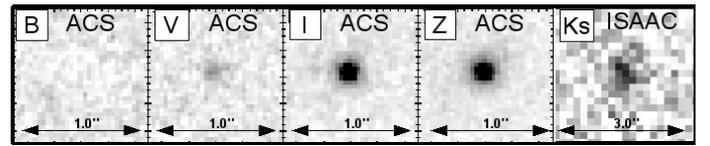}}
\caption{Cutouts of the source GDS~J033218.92-275302.7. From left to right:
the \wb\ , \wv\ , \wi\ , and \wz\ HST/ACS bands. The box side is 1.0 arcsec
(6 kpc proper at the redshift of the source). Drawn from the
V2.0 ACS catalog {\it http://archive.stsci.edu/prepds/goods/}. The last right box
is the ISAAC/$Ks$ band.  The box side is 3 arcsec.}
\label{fig4}
\end{figure}
As shown in Fig.~\ref{fig4}, the source has been detected in the
\wv\ band with a magnitude of 27.63$\pm$0.22, showing an attenuation 
of $\sim$ 94\% to respect the 1400\AA~emission (\wz\ band), which is 
consistent with the average IGM transmission at this redshift \citep[e.g.][]{songaila04}.

Apart from the IGM attenuation that influences the bands bluer than the \wz, 
the main feature of the SED 
is the discontinuity detected between VLT/ISAAC  $J$, $H$, and $Ks$ bands
($\lambda\lesssim 3400\AA$ rest-frame) and the Spitzer/IRAC channels 
($\lambda\gtrsim 5400\AA$ rest-frame), see Fig.~\ref{fig_SED}.
Such a discontinuity is consistent with estimates available in
the literature for the same object
(FIREWORKS, \cite{wuyts08}, \cite{stark07}, \cite{wik08}, \cite{raiter09}). 

The typical uncertainties
(1$\sigma$) span the range 0.03 to 0.11 going from the HST/ACS, VLT/ISAAC
and Spitzer channels 1 and 2, for the last two Spitzer bands 
(5.6 and 8 $\mu m$) the errors increase to $\sim$ 0.2/0.3 mags.

\begin{figure*}[] \centering 
 \includegraphics[width=9cm]{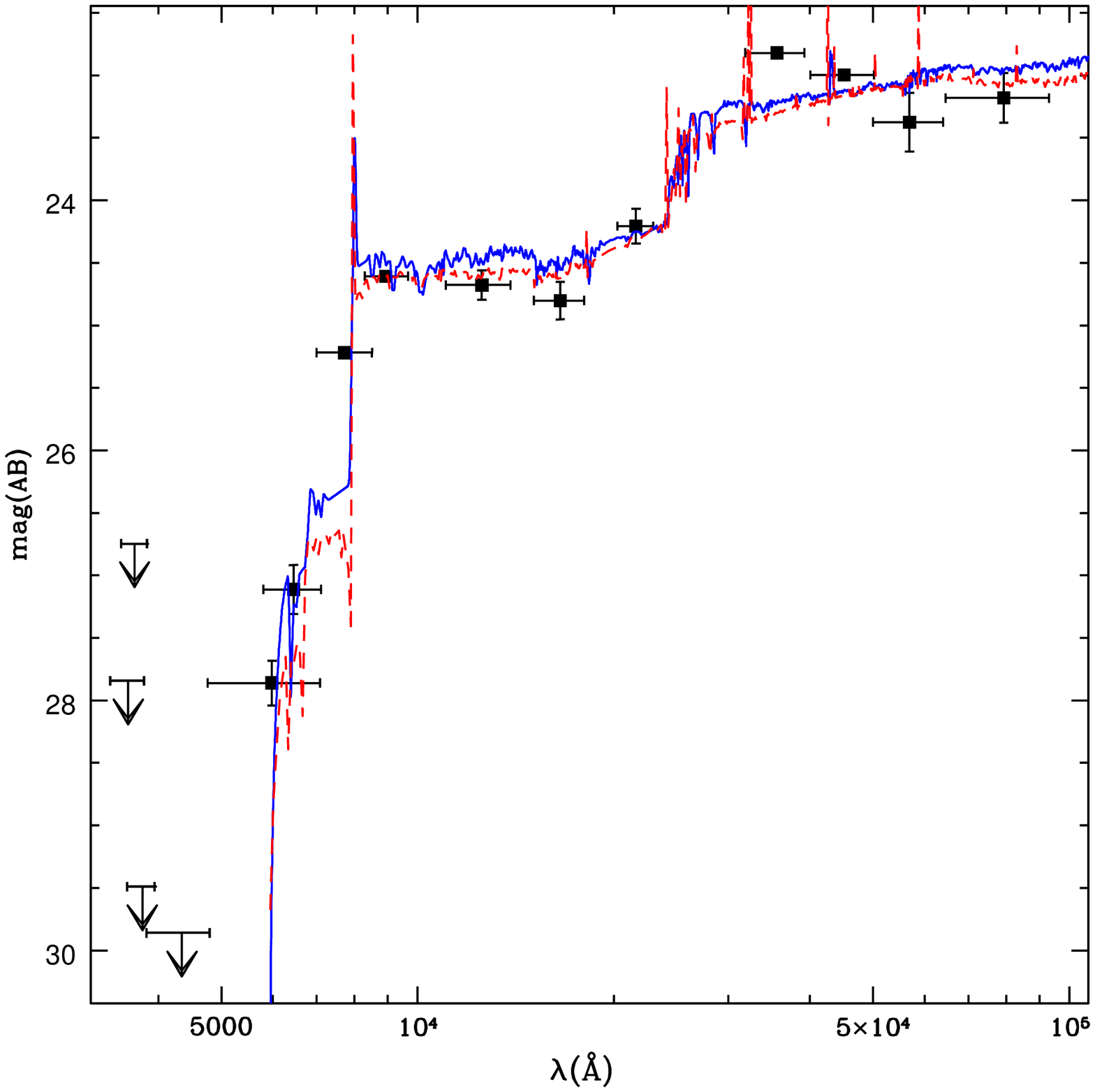}
 \includegraphics[width=9cm]{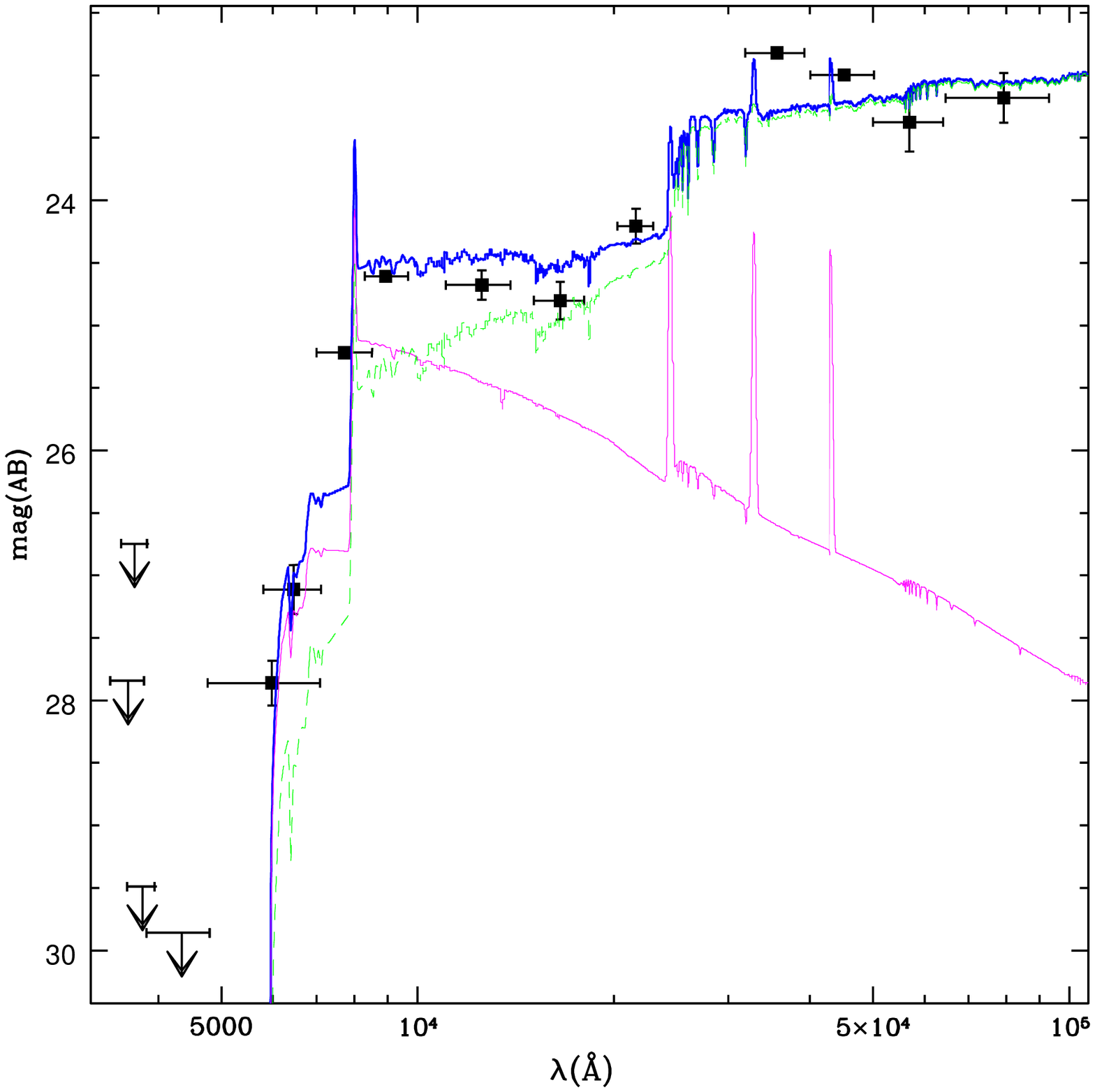} 
\caption{The resulted template fitting over the MUSIC multi-band catalog
of the GDS~J033218.92-275302.7. {\bf Left:} Single stellar population modeling. 
Blue solid line is the fit adopting the maximum ratio [O\,{\sc iii}]/[O\,{\sc ii}] 
(prescription ''Single/[O\,{\sc iii}] max'' in Table~\ref{popgal}),
and red dashed line is the fit with the Schaerer \& de Barros (2009) method
(''Single SB09'' in the same table), see text for details.
{\bf Right:} Double stellar population modeling. Green dashed line shows the 
evolved component (age $\sim$0.4 Gyr), the thin cyan line
 the young contribution (age $\sim$ 0.01 Gyr), and black thick line the best fit 
 composition of the two.}
\label{fig_SED}
\end{figure*}

\begin{table*}
\centering 
\caption{Summary of the photometric information (magnitudes and 1-$\sigma$ errors) for our source.}
\label{tab3}
\begin{tabular}{lccccccccccccc}
\hline\hline
$U_{VIMOS}$$\dag$ &\wb\  & \wv\ & R   & \wi\ & \wz\ & Js  & H   & Ks  &3.6$\mu m$&4.5$\mu m$&5.8$\mu m$&8.0$\mu m$&24$\mu m$ \\
\hline
\hline
29.49        &29.86      &27.86 &27.11&25.22 &24.61       &24.68 &24.80&24.21 &22.82      &23.00     &23.37     &23.18 & 22.61 \\
\hline
1$\sigma$ l.m&1$\sigma$ l.m& 0.179     & 0.195   & 0.039 & 0.031     & 0.116    & 0.151   & 0.140    & 0.014    &  0.024   & 0.235    & 0.198 & 1$\sigma$ l.m.   \\
\hline
\multicolumn{14}{l}
{$\dag$ Other two WFI U bands are available, U35 and U38 with slightly different filter shape. The lower limits
are 27.84 and 26.75, respectively,}\\
\multicolumn{14}{l}
{much shallower than the limit provided by the $U_{VIMOS}$.}\\
\end{tabular}
\end{table*}

\subsection{Morphological properties}

Image cutouts of the isolated source GDS~J033218.92-275302.7 in
the \wb\ , \wv\ , \wi\ , \wz\ (HST/ACS), and $Ks$ (VLT/ISAAC) bands
are shown in Fig.~\ref{fig4}, where each box is 1.0 arcsec wide for the ACS figures 
(6 kpc proper at the redshift of the source), while it is 3 arcsec on a side for the 
ISAAC $Ks$ image.
We considered the ACS/\wz\ and the ISAAC/Ks bands
to derive basic morphological quantities (see Table~\ref{tab1}).
\begin{figure}[] \centering
  \resizebox{\hsize}{!}{\includegraphics{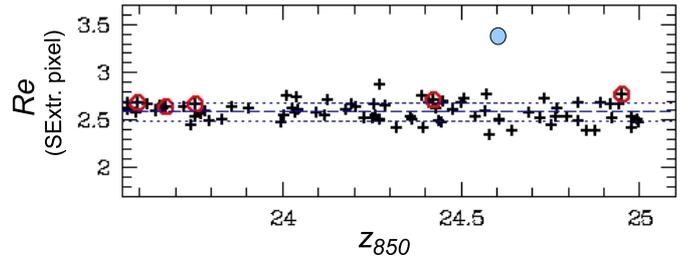}}
\caption{The effective radius measured with the SExtractor algorithm (pixel units) as a function 
of the \wz\ magnitude for a sample of stellar-like sources selected having S/G classifier larger than 0.97 (crosses).
Solid line is the median value and the dotted lines the 1$\sigma$ percentiles.
The open red circles are stars confirmed spectroscopically. The filled blue circle indicates the 
source described in the present work.}
\label{stars}
\end{figure}

\begin{itemize}
\item{{\bf \wz\ band ($\lambda_{rest}\sim$ 1400\AA)}

The uncorrected-PSF effective radius ($Re$) available 
from the ACS/GOODS public catalog v2.0 is 3.37 pixels.
The same quantity derived from a sample of 80 stellar-like sources (SExtractor star/galaxy index larger than 0.97, 
1.0=star, 0.0=extended source) with \wz\ magnitude in the range 23.5-25 gives a median
value and 1$\sigma$ percentiles of $2.590_{-0.098}^{+0.091}$, implying that this source 
is not a stellar-like object (see Fig.~\ref{stars}).
The same result is obtained for the \wi\ band ($\sim$6$\sigma$ from the median value of the stars). 
We note that the SExtractor star/galaxy index of the source is quite high, 0.83, but
lower than the typical value of the stars.
Therefore, even though the present galaxy is clearly a compact source in the UV rest-frame,
it is marginally resolved both in the \wi\ and \wz\ bands.

To derive PSF-corrected morphological parameters, we ran the GALFIT program
(\cite{peng02}) in both bands. The morphological shape of the source is not particularly
complicated so a good fit is reached (reduced $\chi^{2}$ = 0.591) by adopting a simple Gaussian
profile (S\`ersic model with $n$=0.5) 
and leaving the $Re$, the axis ratio $B/A$, the coordinates $X,Y$, the magnitude, and the position angle 
as free parameters (SExtractor estimates were used as a first guesses). 
In the left panel of Fig.~\ref{galfit} the \wz\ band image of the galaxy is shown, and
the residuals map provided by GALFIT, as a result of the subtraction of the best-fit 
model from the original galaxy, do not show significant structures (middle panel of the same figure).
An effective radius $Re$=0.62$\pm$0.05 pixels (0.11$\pm$0.01 kpc physical) and $B/A$=0.61$\pm$0.14 have been obtained. 
We explored how the variation in $Re$ affects the residuals fixing n=0.5 and B/A=0.61. GALFIT was 
run 80 times by varying the $Re$ from 0.025 to 2.0 pixels with a step of 0.025.
Its behavior is shown in the right panel of Fig.~\ref{galfit} where the minimum of the residuals
for $Re\sim$0.650 is clear, still with acceptable values between $Re$ 0.3-1.0 (pixels). 
It is worth noting that, if we perform the
same fit in the \wi\ band, the best $Re$ is slightly smaller than the \wz\ band (0.425 pixels, i.e. 0.013 arcsec or
0.08$\pm$0.01 kpc physical), implying an even more compact region for the \lya\ emission.}

\begin{figure}[] \centering
  \resizebox{\hsize}{!}{\includegraphics{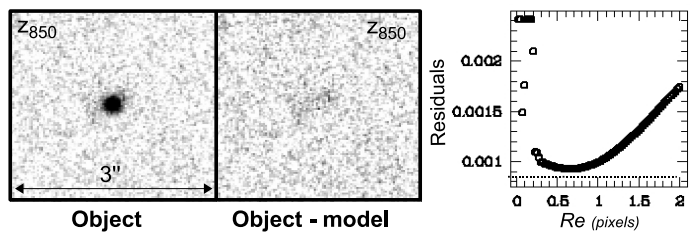}}
\caption{The ACS \wz\ image of the source (left, 100$\times$100 pixels) and the 
residual image after subtraction of the model
derived from GALFIT (middle). 
In the right panel the behavior of the residuals (standard deviation calculated on a 
20$\times$20 pixels area centered on the source) as a function of the $Re$ (fixed during GALFIT runs) is shown.
The minimum value corresponds to $Re\sim$0.65 pixels, i.e. 0.02 arcsec or 0.12 kpc proper. The dotted
line indicates the median value of the background residuals (standard deviation calculated in the blank regions).}
\label{galfit}
\end{figure}

\item{{\bf $Ks$ band ($\lambda_{rest}\sim$ 3300\AA).}

To approach the optical rest-frame
wavelengths, we exploit
the information derived from the near infrared (NIR hereafter) observations (VLT/ISAAC Ks band).
The resolution from the ground is not comparable to that obtained from the
space; however, we note that the present object has an 
FWHM fully consistent with the seeing at the epoch of
observations of that particular region of the GOODS field (0.50 arcsec), i.e., 
it is not resolved. Assuming a Gaussian shape with an FWHM
of 0.50 arcsec, it turns out that the $r_{e}[3300\AA]$ is 0.9 kpc physical. 
For the Gaussian profile the $Re$=(FWHM/2)/1.738, i.e. the radius containing 
half of the total light. Since the source is not resolved, the derived $Re$ is 
an upper limit at this wavelength.}

\end{itemize}

There are three other sources (with secure redshift, quality ``A'') in the FORS2 
sample that show a similar \wz\ UV morphology of the source discussed here
(with a SExtractor $Re$ smaller than 3.5 pixels (0.105 arcsec)).
One is the QSO GDS~J033229.29-275619.5 (\cite{fontanot07}) fully
compatible the stellar PSF
and two other LBGs: GDS~J033217.22-274754.4 at redshift 3.652 already described in \cite{vanz08}
with a double structure of the \lya\ profile and GDS~J033240.38-274431.0 at redshift 4.120,
which shows a continuum with clear \lya\ emission and no other features.

\section{Possible scenarios for GDS~J033218.92-275302.7}

\subsection{A chance superposition ?}

If we interpret the emission line detected at $\lambda$=9742\AA~as a
[O\,{\sc ii}] 3727 foreground emitter superimposed on a background 
and brighter higher-z (z=5.563) LBG, its redshift would be 1.614. 
It is well known that in the GOODS-S
field there is an overdensity structure at that redshift, z=1.61 
\citep[e.g.][]{vanz08,caste07,kurk09}.
However we can exclude this possibility on the basis of the ACS morphology, shows
a compact and circular shape (see Fig.~\ref{fig4}) and from the ultradeep U band observations
carried out by the VLT/VIMOS instrument (\cite{nonino09}), provide an upper limit 
of $\sim$30 AB at 1$\sigma$ (also it has not been detected in the ACS \wb\ band image). 
The source should be detectable in the blue if there is
star formation activity traced by the [O\,{\sc ii}] 3727 line. 
Moreover, assuming a flat continuum at that 1$\sigma$ limit (30 AB), the rest-frame [O\,{\sc ii}] 3727
equivalent width would be larger than 10$^4$\AA. Even though we consider this possibility
largely unlikely, we note that an example of strong [O\,{\sc ii}] 3727 emitter has been 
reported by \cite{stern00}.

\subsection{Is it an AGN ?}

The spectral range up to 10164\AA~allows us to detect
possible emission lines testing for the presence of an AGN
(e.g. N\,{\sc v}] 1240-1243, SiIV 1393.8-1402.8, and C\,{\sc iv}] 1548.2-1550.8). 
Those features are routinely detected in spectra of the most obscured AGNs 
\citep[e.g.][]{polletta06,polletta08}.

\begin{figure}[] \centering
  \resizebox{\hsize}{!}{\includegraphics{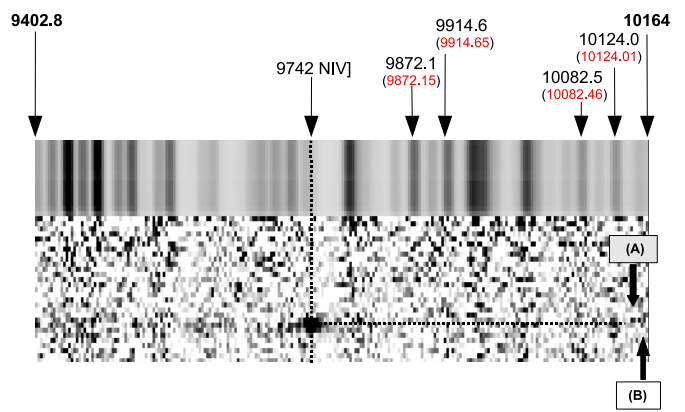}}
\caption{Extracted VLT/FORS2 2-dimensional spectrum of the galaxy discussed 
in the present work (the spectral interval 9403-10164\AA~is shown). 
As a check of the wavelength calibration the skyline position derived
from the FORS2 spectrum are reported (the sky line measurement performed by 
VLT/UVES and Keck/HIRES are indicated within parenthesis 
{\it www.eso.org/observing/dfo/quality/UVES/pipeline/sky$\_$spectrum.html}).
The expected positions of the C\,{\sc iv}] feature are also shown with thick
arrows: case (A) at the redshift of the observed N\,{\sc iv}] 1486 (z=5.553) and case (B) 
at the redshift of the \lya\ line (z=5.563); see text for details.}
\label{fig_end_spec}
\end{figure}

\subsubsection{Line emission in the UV}

\begin{itemize}

\item{{\bf N\,{\sc v} 1240-1243 feature.}

Observationally, the N\,{\sc v} emission is often present in the AGN case.
The FORS2 spectrum allow us to measure emission features down
to $\sim$2$\times$$10^{42}$$erg/s$.
No N\,{\sc v} line is detected (see Fig.~\ref{fig_zoom}).}

\item{{\bf C\,{\sc iv} 1548.2-1550.8 feature.}

Unfortunately, the C\,{\sc iv} line is
at the very red limit of the observed spectrum,
which in this particular slit position is determined by the detector end.
Figures~\ref{fig_end_spec} and \ref{fig_zoom} (top panel) show the 2-dimensional
and the 1-dimensional spectra zoomed at the red edge, respectively.
On one hand, if the redshift of the C\,{\sc iv} feature
is higher than 5.565 (z(C\,{\sc iv})$>$5.565, higher 
than the \lya\ redshift), then the doublet is completely out. 
Alternatively, if z(C\,{\sc iv})$<$5.554, then the doublet is completely in.
In general, LBGs show the observed \lya\ redshift as higher than
the other spectral features because of its asymmetry which 
typically arises from backscattering of the receding 
material \citep[e.g.][]{shapley03,vanz09}. Similarly, in the AGN case, a blueshift 
of the C\,{\sc iv} feature with respect to the observed \lya\ peak is typically observed, 
of several hundred kilometers per second. In particular a blueshift of 
$\sim$ 600$\pm$100 $km~s^{-1}$ with respect to the \lya\ line is measured in the 
SDSS QSO composite spectrum (e.g. see Table 4 and Figure 9 of \cite{vanden01}).
If this is the case, the C\,{\sc iv} feature should fall in
the available spectrum, and its luminosity limit is
$\sim$6$\times$$10^{42}$$erg/s$ at 2$\sigma$ of the background noise fluctuation. }

\end{itemize}

The limits and luminosities estimates on \lya\ , N\,{\sc v}, 
Si\,{\sc iv}, and C\,{\sc iv} are reported in Table~\ref{tab1}.

\subsubsection{X-ray emission}

We further note that this source has not been detected in the X-ray by
the 2 Ms Chandra ultra-deep observations, neither in the MIPS 24$\mu$m by Spitzer 
(with 1$\sigma$ lower limit of 22.61 AB) nor by the 
VLA at 20 cm down to 8 $\mu Jy$ (at 1$\sigma$, \cite{tozzi09}).

A correlation between 2-10 keV X-ray luminosity and [O\,{\sc iii}] 5007 or H$\alpha$
emission line luminosities is observed for galaxies at $z < 1$ 
(e.g., \cite{panessa06}, \cite{silve08}).  Assuming that this correlation also holds at  
higher redshifts, then we may use the X-ray luminosity to derive a constraint on the  
fluxes for AGN-powered emission lines in the IRAC bands.

From the current 2 Ms observations, there 
is no detection at the position of the source (3$\sigma$ limit  
of $\sim$ 3$\times$$10^{43}$ erg/s at 3-13 keV rest-frame, \cite{luo08}).
This limit roughly corresponds to an upper limit for both H$\alpha$ and [O\,{\sc iii}] 5007 luminosities
of $\sim$ $10^{40.5-42.5}$ and $10^{41-43}$ $erg~s^{-1}$, respectively. 
Such values are affected by large uncertainties in the assumed relations (intrinsic scatter) 
and the limit derived from the 2 Ms image. However, if compared to the 
(at least) one magnitude jump between the VLT/ISAAC and Spitzer/IRAC magnitudes, these estimations 
suggest that, besides a line contribution to the IRAC magnitudes, there is also a 
significant contribution from stellar emission beyond 5000\AA~rest-frame, i.e., of a relatively 
evolved stellar population (see Sect. 4). For example a line luminosity of 7$\times$$10^{45}$$erg~s^{-1}$
is needed to boost the 4.5$\mu m$ AB magnitude from 24 to 23 (adopting a bandwidth of 10100\AA , 
\cite{fazio04}).

\subsubsection{A rare class of QSOs: an open possibility}

It is interesting to compare our N\,{\sc iv}] 1486 emitter spectrum with the composite
spectra of quasars available in the literature. This has already been done 
by several authors (e.g. \cite{baldwin03}, \cite{glikman07}, and \cite{jiang08}).
None of the published average quasar spectral templates show
any trace of N\,{\sc iv}] 1486 emission. Nevertheless, focusing the attention on 
this spectral feature, \cite{bentz04} compile a sample of
6650 quasars in the range 1.6$<$z$<$4.1 showing the N\,{\sc iv}] 1486 line 
(other than the N\,{\sc iii}] 1750), and more recently an updated work 
by \cite{jiang08} (on SDSS data release 5) reported that such objects
are $\sim$ 1.1\%~of the total SDSS quasar sample.
They also note that for this small fraction, the N\,{\sc v} 1240 and \lya\ are much 
stronger than the rest of the population. 
We recall that our source does not show the N\,{\sc v} 1240 line.

More interestingly and similar to our findings,
\cite{glikman07} have
discovered two low luminosity QSOs at redshift $\sim$ 4 showing \lya\ , N\,{\sc iv}] 1486 and C\,{\sc iv} 1548-1550 emissions,
but no detection of N\,{\sc v} 1240. In one case the equivalent width of the N\,{\sc iv}] 1486 is larger 
than the C\,{\sc iv} one (240\AA~vs. 91\AA), while it is the opposite for the other (24\AA~vs. 91\AA).
Our source has a luminosity of $M_{145}$=-22.1 (AB) and shows a clear N\,{\sc iv}] 1486 emission 
with an equivalent width of $\sim$ 22\AA~and FWHM $\sim$ 400 $km~s^{-1}$. The \lya\ line shows a
narrow component with a measured FWHM of $\sim$ 600 $km~s^{-1}$. As performed in
\cite{glikman07}, since the blue side of the line profile is absorbed, 
we forced the symmetry in the line by mirroring the
red side of the line profile over the peak wavelength and computed the Gaussian fit.
The narrow-line component increases to $\sim$ 750 $km~s^{-1}$. The broad-line feature
(indicated with a segment in the innner box of Fig.~\ref{fig3}) gives
an FWHM of $\sim$ 3500 $km~s^{-1}$. This would put the source in the QSO regime
(velocity width larger than 1000 $km~s^{-1}$). Therefore, the present source may
be consistent with the interpretation of a low-luminosity quasar in which
the host starburst galaxy is visible (similarly to \cite{glikman07}). 

The study of 
stellar populations of low-luminosity AGNs (e.g. low-luminosity Seyfert galaxies, 
low-ionization nuclear emission line regions, LINERs, and transition-type objects, TOs) 
has been addressed for the local Universe (e.g., \cite{delgado04}), but 
this is still a poorly explored regime at higher redshift.
While it is beyond the scope of the present work to explore the link between the 
coevolution of (circumnuclear) starburst activity and the central black hole accretion,
we simply note that both AGN and star-formation required gas to fuel them, 
and it happens on different temporal and spatial scales, on sub-parsec and typically above few
hundred parsecs (up to several kilo-parsecs) regions, respectively (e.g., \cite{davies07},
\cite{chen09}). In the present case,
the size of the UV emitting region is compact, but
still resolved in the \wz\ ACS image (as shown in Sect. 2).
In summary, the presence of an AGN -- in a rare evolutionary stage -- may be indicated 
by the N\,{\sc iv}] 1486 and broad \lya\ features, even though N\,{\sc v} 1240, Si\,{\sc iv} 1394-1493, 
and (possibly) C\,{\sc iv} 1548-1550 are not detected.

\subsection{A multi-burst galaxy in a peculiar stage of evolution ?}

The source GDS~J033218.92-275302.7 has already been 
analyzed in \cite{wik08}, who classify it as a ``pure'' balmer break galaxy
(their ID $\#5197$). The discontinuity
detected between the $Ks$ and 3.6$\mu m$ bands is interpreted as a signature 
of the Balmer break, suggesting a relatively evolved age of stellar populations with a
significant stellar mass already in place (age of $\sim$0.7Gyr and 
$M^{*}$$\sim$7-8$\times$$10^{10}M_{\odot}$).
A similar conclusion has been reached by \cite{stark07}, 
who find an even higher stellar mass of $10^{11}M_{\odot}$ 
(their ID 32$\_$8020).

However, most probably the observed ($Ks$-3.6$\mu m$) color is
contaminated by emission lines in the 3.6$\mu m$ band, e.g.
[O\,{\sc iii}] 4959-5007. A similar boost to the flux
in the 4.5$\mu m$ band may come from the H$\alpha$ emission line.
It was also selected as H$\alpha$ emitter by R. Chary et al., private
communication.
Apart from the evident \lya\ emission,
which implies the presence of young ($<$ 10 Myr) stars -- i.e., some ``current''/ongoing star formation --
significant nebular emission is also robustly supported by the detection
of the N\,{\sc iv}] 1486 line.
As mentioned above, a similar feature has been identified in the $Lynx$ arc and may
indicate a short powerful starburst in which very hot and massive stars
(T$\gtrsim$80000 K, \cite{fosbu03}) or cooler Wolf-Rayet stars are involved (\cite{VM04}).
A similar blackbody ionizing source may be present in this source.
The ongoing star formation activity would also be
responsible for the measured outflow, whose spectral signature is in the red tail
of the \lya\ profile (see \lya\ modeling in Sect. 4.2).
It is beyond the scope of the present work to model the ionizing source;
nevertheless, we note that in a ``pure'' nebular scenario,  
the continuum is practically flat, and the observed ``breaks'' are produced by
strong nebular emission lines (see \cite{raiter09} for a dedicated discussion).
Alternatively, a different interpretation suggests a contribution from a relatively
evolved stellar population that produces the Balmer break signature (see next section).
Given the current spectroscopic and photometric information,
the following $mixed$ scenario may be possible: 1) ongoing active
star formation in an $H II$-like region that produces nebular emission,
as probed by the \lya\ and N\,{\sc iv}] 1486 features, and 2) an already evolved population of
stars formed at higher redshift, as probed by the signal detected in the IRAC
bands, in particular, redwards of the 4.5$\mu m$ (beyond $\sim$7000\AA~rest-frame).

\section{SED and \lya\ modeling}

We cannot definitively distinguish between the two scenarios described above, in particular 
for the explanation of the N\,{\sc iv}] 1486 feature. In either case, even though the source reflects an early 
stage of coevolution of the (circumnuclear starburst) galaxy 
with its AGN or it is an $HII$ source, the features of the host galaxy are detected 
and can be investigated.
We therefore need to model the SED allowing for multiple
stellar populations. Moreover, valuable information can be derived from the \lya\ line modeling.
This is performed in the following sections.

\subsection{Modeling the SED}
The SED modeling was performed adopting the multiwavelength
GOODS-MUSIC photometric catalog (\cite{grazian06}), and the
spectral fitting technique was developed in 
\cite{fontana03,fontana06} (similar to those adopted in other
works, e.g., \cite{dick03,drory04}).
In the previous section we pointed out that this galaxy likely
contains a mixed stellar population, both young stars, as implied by
a bright Ly$\alpha$ line, the N\,{\sc iv}] 1486 line, and old
stars that produce the 4000\AA~break, clearly observed in broad band
photometry.  We therefore model the SED of the galaxy both with a
single stellar population and with a more plausible double stellar
population. More complicated mixes of multiple stellar populations becomes
unconstrained by the data given the many
degrees of freedom of each population. We actually reduce the 
number of degree of freedom by imposing the requirement that 
both populations (old and young) are 
affected by the same dust extinction (with a Calzetti
or a Small Magellanic Cloud extinction curve).  
Although this might not be true, we feel that it is
plausible for very compact objects such as the one we are studying.  We
fix the rest-frame equivalent width of the \lya\ line to be EW=60\AA, as
measured from the spectrum and regardless of the star formation rate.

As discussed in the previous section, the contribution of nebular
lines to the photometry may affect the IRAC magnitudes of the 3.6$\mu m$
and 4.5$\mu m$ bands.  In particular, the [O\,{\sc iii}] emission contributes to
the 3.6 $\mu m$ channel and the H$\alpha$ line to the 4.5 $\mu m$
channel.  While the H$\alpha$ line can be modeled relatively easily
and its luminosity can be assumed to be proportional to the global SFR
through the well-known Kennicutt relations, disentangling the [O\,{\sc iii}]
contribution is harder.  \cite{moustakas06} investigated the [O\,{\sc iii}] nebular 
emission line as a quantitative SFR diagnostic and
conclude that the large dispersion in the [O\,{\sc iii}]/H$\alpha$ ratio
among star-forming galaxies precluded its suitability for SFR studies.

We therefore treat the [O\,{\sc iii}] contribution in three
  different ways: 1) by assuming a {\it mean} [O\,{\sc iii}] flux as inferred in
  local star-burst galaxies (corresponding to a ratio
  f([O\,{\sc iii}])/f([O\,{\sc ii}])=0.32), 2) by assuming a maximum [O\,{\sc iii}] flux in the 3.6
  $\mu m$ band 10 times larger than in the previous case,
  corresponding to the maximum observed [O\,{\sc iii}]/H$\alpha$ in
  star-burst galaxies, and 3) by neglecting the 3.6 $\mu m$ band in the fit.

 In the SED-fitting computation, the formal errors of the observed magnitudes 
 have a minimum value permitted for each band.
 This was done to avoid over-fitting in the $\chi^{2}$ minimization
 procedure, and it affects only the 3.6 and 4.5 $\mu m$ bands, whose errors
 are increased to 0.1 (the minimum permitted) during the fit.
 
The results of the various fits are reported in Table~\ref{popgal}: for
 each model, we report the best-fit ($bf$) total stellar mass, age, $\tau$ 
 (the star formation e-folding timescale), current SFR, and
  $E(B-V)$ (indicated with $EBV$) for the single and double populations, as well as the minimum
  and maximum values allowed by the fit (at 1$\sigma$).
  For the double stellar populations the best-fit 
  ages of the young and the evolved components are reported (in the last two columns).

\begin{table*}
\centering
\caption[]{SED modeling: results.}
\label{popgal}
\begin{tabular}{lllllllllllll}
\hline
\hline
MODEL       & $M_{min}$ & $M_{bf}$ & $M_{max}$ & $SFR_{min}$ & $SFR_{bf}$ & $SFR_{max}$  & $EBV_{min}$ & $EBV_{bf}$ & $EBV_{max}$           & $T$ & $\tau$\\  
            & $10^{10} M_{\odot}$   &$10^{10} M_{\odot}$   &$10^{10} M_{\odot}$ & $M_{\odot} yr^{-1}$  & $M_{\odot} yr^{-1}$  & $M_{\odot} yr^{-1}$ & & & &Gyrs &Gyrs \\
\hline
Single / [O\,{\sc iii}] ave & 5.2 & 6.8 & 9.1 & 22 & 29 & 30 & 0.00 & 0.00 & 0.00 & 0.71& 0.3\\
Single / [O\,{\sc iii}] max & 5.2 & 6.8 & 9.1 & 22 & 29 & 30 & 0.00 & 0.00 & 0.00 & 0.71& 0.3\\
Single /  no 3.6    & 3.4 & 4.5 & 6.8 & 22 & 34 & 50 & 0.00 & 0.03 & 0.06 & 0.79& 0.6\\
Single SB09 $\dag$ & 4.1 & 6.4 & 8.4 & 17 &36 & 41& 0.00 & 0.05 & 0.10 & $<$1&0.7\\
\hline
Double / [O\,{\sc iii}] ave & 4.0 & 5.7 & 6.9 & 22 & 54 & 249& 0.00 & 0.03 & 0.15 & 0.01,0.4& 0.6,0.1\\
Double / [O\,{\sc iii}] max & 4.2 & 6.2 & 6.9 & 21 & 79 & 145& 0.00 & 0.03 & 0.10 & 0.01,0.4& 0.6,0.1\\
Double / no 3.6     & 3.6 & 5.5 & 6.0 & 23 & 81 & 146& 0.00 & 0.06 & 0.15  &        0.01,0.4& 2.0,0.1\\
\hline
\multicolumn{12}{l}
{$\dag$ SED-fitting according to the Schaerer \& de Barros (2009) method.}
\end{tabular}
\end{table*}
\subsubsection{A single stellar population model}
Although the single stellar population model with a declining exponential
SFR is clearly an oversimplification it can set useful limits.
From Table~\ref{popgal}, we see that in all cases the best-fit stellar mass is well above $10^{10} M_{\odot}$
and the age more than 700 Myrs, implying a formation redshift z$>$13.
The variation in the [O\,{\sc iii}] flux of a factor 10 does not have a strong impact on the values 
of mass, age and SFR. 
Even neglecting the 3.6$\mu m$ band,
the stellar mass is set to  4.5$\times$10$^{10}$$M_{\odot}$ (with a minimum value of 3.4),
but the most notable change is that some dust extinction is allowed, with a best fit  
$E(B-V)_{stars}$=0.06.

It is worth noting that the stellar mass and ages we find are comparable to
those derived by \cite{wik08} and \cite{stark07} if the nebular line treatment
is not inserted. If we insert the [O\,{\sc iii}] prescription, our estimates become
somewhat smaller, even though still significatively large given the redshift of 
the source (corresponding to $\sim$ 0.91 Gyrs after the Big-Bang). 
In particular,  \cite{stark07} derive a mass of   $1.4 \times 10^{11} M_{\odot}$
 but without including the 5.8$\mu m$ and 8$\mu m$ IRAC bands in the SED fitting, 
while Wiklind et al. report a mass of 7$\times 10^{10} M_{\odot}$ but assume a 
photometric redshift of 5.2.

For comparison we have also fitted the SED with the method
of Schaerer \& de Barros (2009) allowing for numerous emission lines. The results obtained
(see Table~\ref{popgal} and Fig.~\ref{fig_SED}) are compatible with the two other approaches used
here (see also next section). In addition, these models predict an intrinsic \lya\ equivalent
width of EW(\lya) $\sim$ 49-70 \AA\ (1 $\sigma$ interval), in good
agreement with the observations. It is worth mentioning that the present source is different 
from those analyzed by Schaerer \& de Barros (2009), extracted from a sample of z$\sim$6
star-forming  galaxies of \cite{eyles07}. 
First, the photometric break between the NIR bands and
the first two Spitzer/IRAC channels 3.6 and 4.5 $\mu m$ is at least twice for our source, 1.5 mags.
The source is brighter in absolute scale, allowing smaller photometric errors see Table~\ref{tab3}.
The main nebular contributors at z$>$5.6 to the IRAC channels 3.6 and 4.5 $\mu m$ are 
the H$\beta$, [O\,{\sc iii}], and H$\alpha$ lines, respectively. In this case the H$\beta$ line falls on the
blue edge of the 3.6 $\mu m$ filter transmission (\cite{fazio04}), so its contribution is further attenuated.
However, more importantly, in our case the source has also been detected where the contribution of
nebular lines is less effective (at this redshift), i.e. in the 5.8 and 8.0 $\mu m$ bands. Even though
the photometric errors slightly increase in these bands, the break is still large ($\sim$ 1 magnitude). 
For comparison, in the sample of \cite{eyles07}, the source SBM03$\#$1 (with \wz\ = 25.35) 
has not been detected in the reddest IRAC channels 3 and 4. This is the main reason for this galaxy 
maintaining a relatively large estimation of the age and stellar mass, 
even considering the nebular contribution.

\subsubsection{A mixed stellar population model}
In this case we allow for two stellar populations with different redshifts 
of formation and different star formation histories. The only constraint we impose 
(a part the spectroscopic redshift) is that both stellar populations are affected by the same dust extinction.
In Table~\ref{popgal} the best-fit values for total stellar mass and the star formation rate corresponding 
to the sum of the two contributions are shown, each taken with the relevant normalization factor, while 
we separately report two best-fit ages for each population.
In all cases, the best-fit solution is made by an older stellar component (of $\sim$ 0.4 Gyrs) that 
contributes to most of the stellar mass, while the young component has high values of SFR.

The global stellar mass remains similar to the previous single population case,
regardless of the [O\,{\sc iii}] treatment,
with mean values of 5-6 $\times 10^{10} M_{\odot}$
and in all cases greater than 4$\times 10^{10} M_{\odot}$.
The SFRs increase by a factor of at least two, and
solutions with values as high as SFR=250$M_{\odot}$$yr^{-1}$ are acceptable.
In all cases, the e-folding timescale of the star formation rate of the young 
population is close to 100 Myrs.
The derived ages of the two stellar populations are 0.01 and 0.4 Gyrs,
independent of the [O\,{\sc iii}] treatment.
The age of the old component is still quite large
(but less than the previous case), implying a formation redshift
around  z$\sim$8. 
Most important, solutions with small but non
negligible dust extinction are always preferred.

Formally the best-fit solution among those including all bands is
the double population with maximum [O\,{\sc iii}] contribution ($\chi^2 =4.1$).
 \footnote{We performed separate
   fits with enlarged errors in the 5.8 and 8$\mu m$ channels to
   explore their possibile underestimation, and the
   consequence is even higher stellar masses, since more
   weight is given to the 3.6 and 4.5$\mu m$ bands.}  
In summary we consider that 5$\times 10^{10} M_\odot$ is a fair estimate of the
total stellar mass and that, considering all uncertainties in the
data and in the modeling, a solid lower limit can be set at
$\sim 3 \times 10^{10} M_\odot$ (the contribution of the young component
to the total stellar mass is negligible, $\sim$ 1\% in all cases). 
The galaxy contains a stellar population that is at least 400 Myrs old, and the
average extinction ($E(B-V)_{stars}$=0.03-0.06) is smaller but not incompatible with the
extinction factor that comes from the \lya\ profile modeling (see below).

\subsection{Modeling of the \lya\ line}

\begin{figure}[h!]
\centering
\includegraphics[width=9cm]{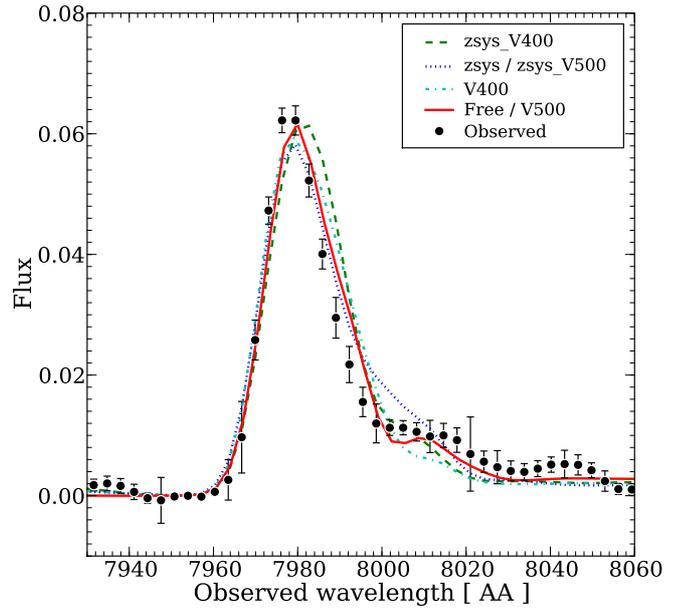}
\caption{The observed \lya\ line profile together with the best-fitting
		synthetic profiles described in the text.}
\label{fig:fitres}%
\end{figure}

\begin{table*} 
\centering
\caption{Best-fit parameters from the \lya\ line fitting. Values marked in bold face have been fixed
 during the fitting procedure.}
\label{tab:fitres}
\renewcommand{\footnoterule}{} 
\begin{tabular}{lllllll}
\hline 
\hline
Param. & {\tt zsys\_V400} & {\tt zsys\_V500}  & {\tt zsys} & {\tt V400} & {\tt V500} & {\tt free} \\
     
\hline
$n(\mathrm{HI})$ [cm$^{-2}$]       & 20.8      & 20.8      & 20.8  & 20.8     & 21.4      &   $21.4^{+0.0}_{-0.6}$  \\
$V_\mathrm{exp}$  [km~s$^{-1}$]     & {\bf 400} & {\bf 500} & 500   & {\bf 400}& {\bf 500} &$500^{+0}_{-201}$\\
$\tau_\mathrm{a}$                  & 3.0       & 3.0        & 3.0  & 3.0       & 3.0       &$3.0^{1.0}_{-2.0}$     \\
$b$              [km~s$^{-1}$]     & 10.0      & 40.0       & 40.0 & 20.0     & 10.0      & $10.0_{-0.0}^{+150.0}$    \\
$z$                             & {\bf 5.553} & {\bf 5.553} & {\bf 5.553}&5.551  & 5.540   & $5.540^{+0.007}_{-0.001}$ \\
$\chi^2_\nu$                    & 3.97         & 3.22       & 3.22  & 3.19    & 1.785   &1.785   \\
\hline
\end{tabular}
\end{table*}

The galaxy shows an \lya\ emission line with an FWHM of
600 $km~s^{-1}$, an evident asymmetric profile, a clear sharp decline in flux on the 
blue side and a red tail of \lya\ photons extending up to $\sim$40\AA~(1500 $km~s^{-1}$)
from the peak of the line (see inner box of Fig.~\ref{fig3} and Fig.~\ref{fig2}).

\lya\ is a resonance line that undergoes a complicated radiation transport, with
the line formation under the influence of numerous parameters: not only 
dust but also the geometry, kinematics, and temperature structure of the 
neutral ISM (e.g. \cite{ahn03}, \cite{ver06}).
These parameters influence the line profile and, 
if sufficient care is taken, the line profile itself can be used to
provide independent and unique constraints (\cite{ver08}). 
Using the Monte Carlo Lyman-alpha ({\em MCLya}) radiation transfer code of
\cite{ver06}, we computed a wide array of possible emergent line
profiles. 

Parameter fitting is performed using a standard least squares fitting engine to
minimize the $\chi^2$ statistic. Details of the
software and fitting can be found in Hayes et al. (2009, in prep).
The parameter space is not entirely unconstrained; e.g., it is possible
to observationally constrain two of the parameters: 
the N\,{\sc iv}] 1486 line puts the systemic redshift at 5.553; and the velocity shift between 
N\,{\sc iv}] 1486 and \lya\ constrain the expanding velocity of the gas shell to respect the 
stellar component, $V_\mathrm{exp}$=457~km~s$^{-1}$ (See Table~\ref{tab1}).
Since our grid of shell parameters is discrete, we adopted the nearest values of the
outflow velocity of 400 and 500~km~s$^{-1}$.
We ran six independent fits in total for all combinations of constraints
denoted as follows: 
{\tt zsys\_V400} and {\tt zsys\_V500} constraining both $z$ and $V_\mathrm{exp}$ (400 and 500 km/s); 
{\tt zsys} constraining $z$; 
{\tt V400} and {\tt V500} constraining $V_\mathrm{exp}$ (400 and 500 km/s, respectively); and 
{\tt free} in which all parameters are fit without constraint.

The results of the fits are presented in Fig.~\ref{fig:fitres} and
Table~\ref{tab:fitres}.
In general, all the fits agree with the case {\tt free}, in particular 
{\tt V500} and {\tt free} produce the same values. 
The \lya\ modeling favors high $HI$ column densities ($N_{HI}>10^{20.8}$ cm$^{-2}$),
outflow velocities of 400-500 km/s (consistent with observations when $V_\mathrm{exp}$
is allowed to vary), and a $\tau_{1216}$$\sim$3.0 for all models 
with $1\sigma$ error of +1/-2 (in the {\tt free} case), which provides a
rough estimate of the extinction $E(B-V)$$\sim$$0.3_{-0.2}^{+0.1}$. 

We also explored the possibility that the emerging \lya\ shape is caused by
a static gas ($V_\mathrm{exp}$=0). In this case the expected double-peaked structure
(e.g. \cite{ver06}) would mimic the single peak observed, since the bluer one could be self-absorbed
(by the galaxy and IGM). From the modeling it turns out that high values of the 
Doppler\footnote{Where the Doppler parameter is $b=(V_{th}^{2}+V_{turb}^{2})^{0.5}$ as the 
contribution of thermal and turbulent motions.}
parameter $b$ and intrinsic FWHM are favored, $\sim$ 160km/s and 1000 km/s, respectively. 
This is not surprising since it is a way to drive photons away from line center in the absence 
of a wind. However, in all cases the resulting fit worsens in general 
and, in particular, the extended red and wavy tail of the line is no longer reproduced. 
Conversely, this feature favor the above interpretation of backscattered 
photons from an expanding shell (the presence of a wind would agree also with the 
ongoing star formation activity).

 It is worth mentioning that we do not expect that the
 extinction undergone by the nebular lines ($E(B-V)_{gas}$ from \lya\ fitting) 
 should match what is undergone by the stars ($E(B-V)_{stars}$, from SED fitting). 
 Indeed Verhamme et al. (2008) find a large scatter in
 the relation between the extinction determined from the
 \lya\ profile fits versus other methods including photometric
 fit and/or measured spectral slopes (see Fig. 12 of their
 work). Calzetti et al. (2001) find an empirical relation of
 $E(B-V)_{stars}$ = 0.44 $E(B-V)_{GAS}$, which would make the two 
 results even more consistent (within their uncertainties).

\section{Discussion}

\subsection{Summary of the modeling}
  
The SED fitting and the \lya\ modeling indicate that:

\begin{enumerate}
\item{The column density of the nebular neutral gas is high, $N_{HI}$$>$$10^{20.8}$. 
 We note that this value is comparable to those found for the damped Lyman-alpha systems, e.g.,  
 \cite{wolfe05}}.
\item{The outflow velocity derived from the \lya\ modeling 
is consistent with the observed one, and it is relatively high (greater than 400 km/s).}
\item{A young and an evolved stellar population are both present. The first with an SFR in the range
30-200 $M_{\odot}$$yr^{-1}$ and  negligible contribution to the total stellar mass (1\%). The 
second with a stellar mass of $\sim$ 5$\times$10$^{10}$$M_{\odot}$ and an age of 0.4 Gyr.}
\item{The extinctions derived from the different methods are compatible
within the 1-$\sigma$ uncertainties and in general are relatively small (in the range 0$<$$E(B-V)$$<$0.3).}
\end{enumerate}

Although the signal-to-noise ratios are low,
the galaxy is strongly detected in both the IRAC 5.8 and 8.0 $\mu m$ bands,
where no nebular lines or strong nebular continuum should contribute
at $z = 5.56$.   Even if emission lines contribute flux to the shorter wavelength
IRAC channels, the longer-wavelength IRAC channels indicate a significant
increase in flux density relative to the rest-frame UV continuum, suggesting the
presence of a Balmer break from an evolved stellar population.

Summarizing, this galaxy shows several interesting 
observed properties: 1) its stellar mass is still relatively high 
($M_{\star}$ $\sim$ 5$\times$10$^{10}$$M_{\odot}$) with a component of already evolved stellar 
populations ($\sim$0.4 Gyr); 2) it contains a star-forming component able to produce
nebular emission lines and with an age of $\sim$ 10 Myr; 
3) a substantial wind is measured both from the optical spectrum and from the \lya\ modeling, 
of 450/500 km/s; and 4) the source is compact in the rest-frame UV and U-band rest-frame
wavelengths.

\subsection{An already dense galaxy ?}

As described in the previous sections, the SED fitting analysis  
implies a stellar mass of $M_{\star}$$\sim$5$\times$$10^{10}$$M_{\odot}$, with a significant, 
evolved component with an age of $\sim$0.4 Gyr. If the very compact size measured in the ACS and ISAAC 
images for the  rest-frame ultraviolet light can be assumed to apply to the overall 
distribution of the evolved stellar population, then it implies a very high stellar 
mass density:

\begin{enumerate}
\item{If we assume a constant size over all wavelengths from
the 1400\AA~to optical bands rest-frame (ACS \wz\ band, 0.11 kpc physical), the stellar density in a spherical symmetric 
shape turns out to be $\rho_{\star}$=(0.5$M_{\star}$)/(4/3$\pi$$r_{e}^{3}$) $\sim$ 3.5$\times$$10^{12}$$M_{\odot}$$kpc^{-3}$.}

\item{Similarly, assuming a constant size over all wavelengths  from the 3300\AA~rest 
frame (ISAAC $Ks$ band, 0.9 kpc physical), the stellar density in a spherical symmetric shape
turns out to be $\rho_{\star}$=(0.5$M_{\star}$)/(4/3$\pi$$r_{e}^{3}$) $\sim$ 8.2$\times$$10^{9}$$M_{\odot}$$kpc^{-3}$.}

\end{enumerate}

In both cases, given the estimations of the effective radius 
and the stellar mass of $\sim$ 5$\times10^{10}M_{\odot}$, 
the source appears to be ultradense if compared with the local
mass-size relation. In case (1) the stellar mass density should be 
considered an upper limit if the $r_{e}$ derived from the 1400\AA~rest-frame is a 
fair estimate of the smallest size.
Locally, on average, the $r_{e}$ is larger than 2 kpc (3 kpc) for early type (late type) galaxies with the 
comparable stellar mass (\cite{shen03}).

Interestingly, the \lya\ modeling suggests (in all cases) a relatively high column density of the neutral gas, 
the $N_{HI}$ turns out in the range $10^{20.8-21.4}$$cm^{-2}$. 
We further note that, assuming that the Schmidt law 
is valid at this redshift (Kennicutt 1998), adopting the observed area of 8.8 $kpc^{2}$
and two possible SFRs estimates (see Table~\ref{popgal}), 30  and 100 $M_{\odot}yr^{-1}$,
the mass of the gas turns out to be $8\times10^{9}$ and $2\times10^{10}$ $M_{\odot}$,
respectively, which represents a significant fraction if compared to the stellar mass
($\sim$ 5$\times$$10^{10}$$M_{\odot}$).

As noted by \cite{bui08}, massive ultradense spheroid observed at intermediate 
redshift $\sim$ 1.5-3 and the globular clusters have remarkably similar stellar densities
(above $10^{10}$$M_{\odot}$$kpc^{-3}$), 
suggesting a similar origin. A massive ultradense galaxy at $z$$\sim$1.5-3 should form
its stars very quickly in earlier epochs and in a high gas-density environment. 
In this sense the present source may represent a ``precursor'' of the ultradense 
spheroids recently discovered at redshift 1.5-3.

\subsection{Feedback in action ?}
The current burst of star formation may be caused by a previous infall 
of gas and/or to a merger event (even though the UV morphology is quite regular).
A vigorous wind of $\sim$ 450 km/s is detected both from the observations 
(\lya\ and N\,{\sc iv}] 1486 velocity offset) and from the \lya\ modeling (leaving all parameters free). 
As discussed above, the \lya\ emission arise from a compact region with an
effective radius not larger than 0.1 kpc (the PSF-deconvolved $Re$ in the \wi\ band is 
$\sim$ 0.08 kpc physical), a possible indication that the outflow of gas is in its
initial phase of expansion from the central region.
This particular phase of the galaxy evolution showing hot and massive stars and/or 
a low-luminosity AGN may be an efficient mechanism to blow the material out from
the potential well of the galaxy, in some way influencing the subsequent star formation
activity and/or the surrounding IGM. 

Wind propagation and escape is quite sensitive to the entrainment fraction 
and to the velocity of the wind itself. This occurs because the two primary
forces limiting wind propagation are the galaxy's potential well and the
ram pressure of the gas that must be swept up even if the wind is fast.
Moreover, if entrainment is significant, then the mass over which the wind
energy and momentum must be shared may be much greater.

Therefore it is first useful to compare the escape velocity from the halo with
the estimated wind velocity. Following the calculation of 
\cite{ferrara00}, the escape velocity can be expressed as

\begin{equation}
 v_{e}^{2} = \frac{2pGM_{H}}{r_{H}}
\end{equation}
 
with $p$=1.65. The isothermal halo density profile is assumed ($\rho_{H}(r)=\rho_{c}/[1+(r/r_{a})^{2}]$),
with an extension out to a radius $r_{200}$=$r_{H}$=$[3M_{H}/4\pi(200\rho_{crit})]^{1/3}$,
defined as the radius within which the mean dark matter density is 200 times
the critical density $\rho_{crit}=3H_{0}^{2}(1+z)^{3}/8\pi G$ at redshift $z$ of
the galaxy. The $r_{H}$ turns out to be $\sim$30 kpc at this redshift, assuming a
halo mass of $M_{H}=10^{12}$$M_{\odot}$.
Under these assumptions, the escape velocity is $v_{e}$ $\simeq$ $540kms^{-1}$, about the same
(or a bit larger) as the wind estimated velocity from spectral features, $\sim$$450kms^{-1}$
(or from the \lya\ modeling, $\sim$$500kms^{-1}$).
It is possible that we are observing the transport of material from the 
galaxy to the halo, which will remain confined. If we assume a slightly lower mass of the halo, e.g.
$M_{H}=5\times10^{11}$$M_{\odot}$ (a factor 10 higher than the stellar mass), then 
the escape velocity turns out to be $v_{e}$ $\simeq$ $380kms^{-1}$
and an $r_{H}$=24 kpc. In this case, the velocity of the wind would be sufficient to
escape the potential well of the halo.

Indeed, from SPH simulations it appears that the main contributors
to the metal enrichment of the low-density regions of the IGM are ``small''
galaxies with stellar masses below $10^{10}$$M_{\odot}$ (\cite{aguirre01b}),
and similar results have been obtained by other authors (e.g. \cite{oppen08}, 
\cite{bertone05}). In the present case the uncertainty on the halo mass
prevents any clear conclusion. If we assume a value lower than 
$10^{12}$ $M_{\odot}$, then the expanding material may reach characteristic distances
(namely ``stall radius'') where the outflow ram pressure is balance by the IGM pressure
up to few hundred kpcs (e.g. \cite{aguirre01a}).

\section{Concluding remarks}

A peculiar galaxy belonging to the GOODS-S field has been discussed.
The main observed features are the relatively strong nebular emission in the 
ultraviolet (\lya\ and N\,{\sc iv}] 1486) and the presence of the 
$Balmer$ $Break$ detected through the NIR VLT/ISAAC and Spitzer/IRAC data.
Indeed, from the SED fitting with single and double stellar populations and the \lya\
modeling, it turns out that the source seems to have an evolved
component with stellar mass of $\sim$5$\times$10$^{10}$ $M_{\odot}$ and age
$\sim$ 0.4 Gyrs, a young component with an age of $\sim$ 0.01 Gyrs 
(contributing to $\sim$ 1\% of the total stellar mass), and 
a star formation rate in the range of 30-200 $M_{\odot}yr^{-1}$.
At present no evidence of common ``N\,{\sc iv}] emitters'' is observed in surveys of
high redshift galaxies or quasars. However, there are rare cases in the literature
that show this line emission (together with other atomic transitions), spanning 
from a pure $HII$ region source to a subclass of low-luminosity quasars. 
In the first case, very hot and massive stars with low metallicity are required to produce
the N\,{\sc iv}] line; however, it is difficult to reproduce the signal
measured in the Spitzer/IRAC channels with a pure $HII$ nebula, in particular 
at wavelengths beyond 4.5 $\mu m$, i.e. 
to reconcile the two observed facts: 1) the presence of a relatively evolved stellar population
and 2) the low-metallicity environment needed if the N\,{\sc iv}] emission arises from stellar
photoionization.
Alternatively, the low-luminosity quasar/AGN interpetation may explain 
the N\,{\sc iv}] emission, the broad \lya\ component, and the properties of the 
host galaxy discussed here, i.e., starforming, massive, and evolved galaxy.

The limits on the size derived from the ACS/\wz\ and VLT/Ks bands 
indicate that this object is denser than the local ones with similar mass, 
with a significant mass of the gas still in place (comparable to the stellar one).
A relatively high nebular gas column density is also favored from the \lya\ line 
modeling, $N_{HI}$$\gtrsim$$10^{21}$$cm^{-2}$, comparable to those found for the
damped Lyman-alpha systems.
The region emitting \lya\ photons is spatially compact,
close to that at the continuum emission at 1400\AA, $\sim$ 0.1 kpc, in which
a vigorous outflow ($\sim$ 450/500 km/s) has been measured from the spectrum 
and \lya\ modeling. 
The gas is expanding from this region, but given the uncertainty on the halo mass,
it is dubious whether it will pollute the IGM to great distances.

Such special objects are the key to understanding fundamental passages in the 
formation and evolution of the galaxy population.
Future instruments will shed light on the nature of this interesting object,
in particular, the JWST and the ELTs will give better and new constraints on the 
optical rest-frame morphology and nebular emission.

\begin{acknowledgements}
 We would like to thank the anonymous referee for very constructive comments and
 suggestions.
 We are grateful to the ESO staff in Paranal and Garching, who greatly helped
 in the development of this program.
 We thank J. Retzlaff for the informations about the VLT/Ks images of the
 GOODS-S field and the useful comments and discussions of
 P. Tozzi, F. Calura, R. Chary, S. Recchi, P. Monaco, and F. Fontanot about the work.
 EV would like to thank Anna Raiter and R.A.E. Fosbury for precious discussions about the photoionization
 modeling. We acknowledge financial contributions from contract ASI/COFIN I/016/07/0 and 
 PRIN INAF 2007 ``A Deep VLT and LBT view of the Early Universe''.
 
\end{acknowledgements}


\begin{thebibliography}{}

\bibitem[Ando et al. (2006)]{ando06} Ando, M., Ohta, K., Iwata, I., Akiyama, M.,
 et al., 2006, \apj, 645, 9

\bibitem[Ando et al. (2007)]{ando07} Ando, Masataka, Ohta, Kouji, Iwata, Ikuru,
Akiyama, Masayuki, Aoki, Kentaro, Tamura, Naoyuki, 2007, PASJ, 59, 717

\bibitem[Aguirre et al. (2001b)]{aguirre01b}
Aguirre, Anthony, Hernquist, Lars, Schaye, Joop, Weinberg, David H., 2001, ApJ, 560, 599A

\bibitem[Aguirre et al. (2001a)]{aguirre01a}
Aguirre, Anthony, Hernquist, Lars, Schaye, Joop, Katz, Neal, Weinberg, David H., Gardner, Jeffrey, 2001, \apj, 561, 521A
	
\bibitem[Ahn et al. (2003)]{ahn03}
Ahn, Sang-Hyeon, Lee, Hee-Won, Lee, Hyung Mok, 2003, MNRAS, 340, 863A

\bibitem[Baldwin et al. (2003)]{baldwin03}
Baldwin, J. A., Hamann, F., Korista, K. T., et al., 2003, \apj, 583, 649

\bibitem[Balestra et al. (2010)]{balestra10}
Balestra, I., Mainieri, V., Popesso, P., Dickinson, M., et al., 2010, A\&A, 512, 12

\bibitem[Beckwith et al. (2006)]{beck06}
Beckwith, S. V. W., et al. 2006, AJ, 132, 1729

\bibitem[Bentz et al. (2004)]{bentz04}
Bentz, M. C., Hall, P. B., \& Osmer, P. S., 2004, AJ, 128, 561

\bibitem[Bertone et al. (2005)]{bertone05}
Bertone, Serena, Stoehr, Felix, White, Simon D. M., 2005, MNRAS, 359, 1201B

\bibitem[Binette et al. (2003)]{binette03}
Binette, L., Groves, B., Villar-Martin, M., Fosbury, R. A. E., Axon, D. J., 2003, A\&A, 405, 975B

\bibitem[Bouwens et al. (2006)]{bouwens06} Bouwens, R.,J., Illingworth, G.,D.,
 Blakeslee, J.,P., Franx, M., 2006, ApJ, 653, 53

\bibitem[Bouwens et al. (2007)]{bouwens07} Bouwens, R. J., 
 Illingworth, G. D., Franx, Marijn, Ford, Holland, 2007, ApJ, 670, 928

\bibitem[Bouwens et al. (2008)]{bouwens08}
Bouwens, R. J., Illingworth, G. D., Franx, M., Ford, H., 2008, \apj, 686, 230

\bibitem[Buitrago et al. (2008)]{bui08} Buitrago, F., Trujillo, I., Conselice, C. J.,
Bouwens, R., J., Dickinson, M., Yan, H., 2008, \apj, 687, 61

\bibitem[Castellano et al. (2007)]{caste07}
Castellano, M., Salimbeni, S., Trevese, D., Grazian, A., 
Pentericci, L., et al., 2007, \apj, 671, 1497C

\bibitem[Cimatti et al. (2008)]{cimatti08} 
Cimatti, A., Cassata, P., Pozzetti, L., Kurk, J., Mignoli, M., 
Renzini, A., et al.,  2008, A\&A, 482, 21C

\bibitem[Conselice et al. (2008)]{conse08} 
Conselice, Christopher J., Rajgor, Sheena, Myers, Robert, 2008, MNRAS, 386, 909

\bibitem[Cristiani et al. (2004)]{crist04}
Cristiani, S., Alexander, D. M., Bauer, F., Brandt, W. N., et al., 2004, \apj, 600, 119

\bibitem[Chen et al. (2009)]{chen09}
Chen, Yan-Mei, Wang, Jian-Min, Yan, Chang-Shuo, Hu, Chen, Zhang, Shu, 2009, \apj, 695, 130

\bibitem[Daddi et al. (2005)]{daddi05}
Daddi, E., Renzini, A., Pirzkal, N., Cimatti, A., et al., 2005, ApJ, 626, 680D

\bibitem[Davies et al. (2007)]{davies07}
Davies, R. I., Sanchez, F. Mueller, Genzel, R., et al., 2007, 2007, \apj, 671, 1388

\bibitem[Dickinson et al. (2004)]{dick04} Dickinson, M., Stern, D.,
Giavalisco, M., Ferguson, H. C., Tsvetanov, Z., et al., 2004, ApJ, 600, 99

\bibitem[Dickinson et al. (2003)]{dick03}
Dickinson, Mark, Papovich, Casey, Ferguson, Henry C., Budavári, Tam\`as, 2003, \apj, 587, 25D

\bibitem[Dickinson et al. (2003b)]{dick03b} Dickinson et al. 2003, in the
proceedings of the ESO/USM Workshop "The Mass of Galaxies at Low and High
Redshift" (Venice, Italy, October 2001), eds. R. Bender and A.  Renzini,
astro-ph/0204213

\bibitem[Drory et al. (2004)]{drory04}	
Drory, N., Bender, R., Hopp, U., 2004, \apj, 616, 103

\bibitem[Elbaz et al. (2007)]{elbaz07}
Elbaz, D., Daddi, E., Le Borgne, D., Dickinson, M., Alexander, D. M., et al., 2007, A\&A, 468, 33E

\bibitem[Eyles et al. (2005)]{eyles05}
Eyles, L. P., Bunker, A. J., Stanway, E. R., Lacy, M., Ellis, R. S., Doherty, M., 2005, MNRAS, 364, 443

\bibitem[Eyles et al. (2007)]{eyles07}
Eyles, L. P., Bunker, A. J., Ellis, R. S., Lacy, M., Stanway, E. R., Stark, D. P., Chiu, K.,  
2007, MNRAS, 374, 910

\bibitem[Fan et al. (2003)]{fan03}
Fan, X., Strauss, M. A., Schneider, D. P., \& Becker. R. H., 2003, AJ, 125, 1649

\bibitem[Fazio et al. (2004)]{fazio04}
Fazio, G. G., Hora, J. L., Allen, L. E., et al., 2004, ApJS, 154, 10F

\bibitem[Ferguson et al. (2004)]{ferg04} Ferguson, Henry C., 
Dickinson, Mark, Giavalisco, Mauro, Kretchmer, Claudia, et al., 2004, ApJ, 600,107

\bibitem[Ferrara et al. (2000)]{ferrara00}
Ferrara, Andrea, Pettini, Max, Shchekinov, Yuri, 2000, MNRAS, 319, 539F

\bibitem[Fontana et al. (2003)]{fontana03}
Fontana, A., Donnarumma, I., Vanzella, E., Giallongo, E., Menci, N., et al., 2003, ApJ, 594, 9

\bibitem[Fontana et al. (2006)]{fontana06}
Fontana, A., Salimbeni, S., Grazian, A., Giallongo, E., Pentericci, L., et al., 2006, A\&A, 459, 745F

\bibitem[Fontana et al. (2009)]{fontana09}
Fontana, A., Santini, P., Grazian, A., Pentericci, L., Fiore, F., et al., 2009, (arXiv/0901.2898)

\bibitem[Fontanot et al. (2007)]{fontanot07}
Fontanot, F., Cristiani, S., Monaco, P., Nonino, M., Vanzella, E., Brandt, W. N., Grazian, A.,
Mao, J., 2007, A\&A, 461, 39

\bibitem[Fosbury et al.(2003)]{fosbu03}
Fosbury, R. A. E., Villar-Martín, M., Humphrey, A., Lombardi, M., et al., 2003, \apj, 596, 797

\bibitem[Giavalisco et al.(2004a)]{Giava04a} Giavalisco, M., Ferguson, H. C.,
Koekemoer, A. M., Dickinson, M., Alexander, D. M., Bauer, F. E., Bergeron, J.,
et al. 2004, ApJ, 600, L93

\bibitem[Giavalisco et al.(2004b)]{giava04b} Giavalisco, M., Dickinson, M.,
Ferguson, H. C., Ravindranath, S., Kretchmer, C., Moustakas, L. A., Madau, P.,
Fall, S. M., Gardner, Jonathan P., Livio, M., Papovich, C., Renzini, A.,
Spinrad, H., Stern, D., Riess, A., 2004, ApJ, 600, 103

\bibitem[Glikman et al. (2007)]{glikman07}
Glikman, Eilat, Djorgovski, S. G., Stern, Daniel, Bogosavljevic, Milan, Mahabal, Ashish, 2007, \apj, 663, 73

\bibitem[Gonz\'alez Delgado et al. (2004)]{delgado04}
Gonzalez Delgado, Rosa M., Cid Fernandes, Roberto, Perez, Enrique, et al., 2004, \apj, 605, 127

\bibitem[Grazian et al.(2006)]{grazian06} Grazian, A., Fontana, A., de Santis, C., 
Nonino, M., Salimbeni, S., Giallongo, E., Cristiani, S., Gallozzi, S., 
Vanzella, E., 2006, A\&A, 449, 951

\bibitem[Grazian et al.(2007)]{grazian07}
Grazian, A., Salimbeni, S., Pentericci, L., Fontana, A., Nonino, M., Vanzella, E., et al.,  2007, A\&A, 465, 393G

\bibitem[Jiang et al. (2008)]{jiang08}
Jiang, Linhua, Fan, Xiaohui, Vestergaard, M., 2008, \apj, 679, 962

\bibitem[Keenan et al. (1995)]{keenan95}
Keenan, F. P., Ramsbottom, C. A., Bell, K. L., Berrington, K. A., et al., 1995, \apj, 438, 500

\bibitem[Kurk et al.(2009)]{kurk09}
Kurk, J., Cimatti, A., Zamorani, G., Halliday, C., Mignoli, M., Pozzetti, L., 2009, A\&A accepted, (arXiv/0906.4489K)

\bibitem[Labb\'e et al. (2006)]{labbe06}
Labb\'e, I., Bouwens, R., Illingworth, G. D., Franx, M., 2006, \apj, 649, 67

\bibitem[Le F\`evre et al. (2005)]{fevre05}
Le F\`evre, O., Vettolani, G., Garilli, B., Tresse, L., et al., 2005, A\&A, 439, 845L

\bibitem[Luo et al.(2008)]{luo08}
Luo, B., Bauer, F. E., Brandt, W. N., Alexander, D. M., Lehmer, B. D., et al., 2008, ApJS, 179, 19L

\bibitem[Miller et al. (2008)]{miller08}	
Miller, Neal A., Fomalont, Edward B., Kellermann, Kenneth I., 
Mainieri, Vincenzo, et al., 2008, ApJS, 179, 114M 

\bibitem[Moustakas et al.(2006)]{moustakas06}
Moustakas, John, Kennicutt, Robert C., Jr., Tremonti, Christy A., 2006, \apj, 642, 775M

\bibitem[Nagao et al.(2008)]{nagao08}
Nagao, Tohru, Sasaki, Shunji S., Maiolino, Roberto, Grady, Celestine, 2008, \apj, 680, 100N

\bibitem[Nonino et al.(2009)]{nonino09}
Nonino, M., Dickinson, M., Rosati, P., Grazian, A., Reddy, N., 
Cristiani, S., et al., 2009, \apj, 183, 244

\bibitem[Oppenheime \& Dav\`e (2008)]{oppen08}
Oppenheimer, Benjamin D., Dav\`e, Romeel, 2008, MNRAS, 387, 577O

\bibitem[Ouchi et al. (2008)]{ouchi08}
Ouchi, Masami, Shimasaku, Kazuhiro, Akiyama, Masayuki, et al., 2008, \apj, 176, 301

\bibitem[Panessa et al.(2006)]{panessa06}
Panessa, F., Bassani, L., Cappi, M., Dadina, M., Barcons, X., et al., 2006, A\&A, 455, 173P

\bibitem[Peng et al. (2002)]{peng02}
Peng, Chien Y., Ho, Luis C., Impey, Chris D., Rix, Hans-Walter, 2002, AJ, 124, 266

\bibitem[Pentericci et al.(2007)]{pente07} Pentericci, L., Grazian, A.,
  Fontana, A., Salimbeni, S., Santini, P., de Santis, C., Gallozzi, S.,
  Giallongo, E., 2007, A\&A, 471, 433

\bibitem[Pentericci et al.(2009)]{pente09}
Pentericci, L., Grazian, G., Fontana, A., Castellano, M., Giallongo, E., 
Salimbeni, S., Santini, P., A\&A, 2009, A\&A, 494, 553P

\bibitem[Polletta et al.(2006)]{polletta06}
Polletta, Maria del Carmen, Wilkes, Belinda J., Siana, Brian, et al., 2006, \apj, 642, 673

\bibitem[Polletta et al.(2008)]{polletta08}
Polletta, M., Omont, A., Berta, S., Bergeron, J., Stalin, C. S., Petitjean, P., 2008, A\&A, 492, 81P

\bibitem[Popesso et al. (2008)]{popesso08}
Popesso, P., Dickinson, M., Nonino, M., Vanzella, E., Daddi, E., et al., 2009, A\&A, 494, 443P

\bibitem[Raiter et al.(2009)]{raiter09}
Raiter, A., R. A. E., Fosbury, H., Teimoorinia and P. Rosati, 2009, A\&A, 510, 109

\bibitem[Ravindranath et al.(2006)]{ravi06}
Ravindranath, Swara, Giavalisco, Mauro, Ferguson, Henry C., Conselice, Christopher,
Katz, Neal, et al.,  2006, ApJ, 652, 963

\bibitem[Santini et al.(2009)]{santini09}
Santini, P., Fontana, A., Grazian, A., Salimbeni, S., Fiore, F., 
Fontanot, F., 2009, A\&A accepted, (arXiv/0905.0683)

\bibitem[Schaerer \& Verhamme(2008)]{sch08} Schaerer, D., Verhamme, A.,
2008, A\&A, 480, 369

\bibitem[Schaerer \& de Barros(2009)]{sch09}
Schaerer, Daniel, de Barros, Stephane, 2009, A\&A accepted, (arXiv/0905.0866)

\bibitem[Shapley et al.(2003)]{shapley03} Shapley, A.E., Steidel, C.C.,
Pettini, M., Adelberger, K.L., 2003, \apj, 588, 65

\bibitem[Shen et al.(2003)]{shen03}
Shen, Shiyin, Mo, H. J., White, Simon D. M., Blanton, Michael R., 2003, MNRAS, 343, 978S

\bibitem[Silverman et al. (2008)]{silve08}
Silverman, J. D., Lamareille, F., Maier, C., et al., 2008, \apj, 696, 396

\bibitem[Songaila (2004)]{songaila04}
Songaila, Antoinette, 2004, AJ, 127, 2598

\bibitem[Stark et al.(2007)]{stark07}
Stark, D. P., Bunker, A. J., Ellis, R. S., Eyles, L. P., Lacy, M., 2007, ApJ, 659, 84S

\bibitem[Stark et al.(2009)]{stark09}
Stark, D. P., Ellis, R. S., Bunker, A., Bundy, K., Targett, T., Benson, A., Lacy, M., 2009, 
\apj, 697, 1493

\bibitem[Steidel et al.(1999)]{Ste99} 
Steidel, C.C., Adelberger, K.L., Giavalisco, M., Dickinson, M., Pettini, M., 1999, ApJ, 519, 1

\bibitem[Stern et al. (2000)]{stern00}
Stern, D., Bunker, A., Spinrad, H., Dey, A., 2000, \apj, 537, 73 

\bibitem[Szokoly et al. (2004)]{szo04}
Szokoly, G., P., Bergeron, J., Hasinger, G., Lehmann, I., Kewley, L., Mainieri, V., Nonino, M., Rosati, P.,
Giacconi, R., Gilli, R., Gilmozzi, R., Norman, C., Romaniello, M., Schreier, E., Tozzi, P., Wang, J., X.,
Zheng, W., Zirm, A., 2004, ApJS, 155, 271

\bibitem[Taniguchi et al.(2005)]{tanigu05} Taniguchi, Y., Ajiki, M, Nagao, T.k,
et al., 2005, PASJ, 57, 165

\bibitem[Taniguchi et al.(2009)]{tanigu09}	
Taniguchi, Y., Murayama, T., Scoville, N. Z., Sasaki, S. S., et al. 2009, (arXiv/0906.1873)

\bibitem[Tapken et al.(2007)]{tapken07} Tapken, C., Appenzeller, I., Noll, 
S., Richling, S., Heidt, J., Meinkohn, E. and Mehlert, D., 2007,A\&A, 467, 63

\bibitem[Taylor-Mager et al.(2007)]{TM07}
Taylor-Mager, Violet A., Conselice, Christopher J., Windhorst, Rogier A., Jansen, Rolf A., 2007, \apj, 659, 162

\bibitem[Tozzi et al. (2009)]{tozzi09}
Tozzi, P., Mainieri, V., Rosati, P., Padovani, P., et al., 2009, \apj, 698, 740

\bibitem[Trujillo et al.(2007)]{trujillo07}
Trujillo, Ignacio, Conselice, C. J., Bundy, Kevin, Cooper, M. C., 2007, MNRAS, 382, 109T

\bibitem[Vanden Berk et al. (2001)]{vanden01}
Vanden Berk, Daniel E., Richards, Gordon T., Bauer, Amanda, et al., 2001, AJ, 122, 549

\bibitem[van Dokkum et al.(2008)]{dokkum08} 
van Dokkum, Pieter G., Franx, Marijn, Kriek, Mariska, Holden, Bradford, 
Illingworth, Garth D., Magee, Daniel, et al., 2008, \apj, 677, 5

\bibitem[Vanzella et al.(2005)]{vanz05} Vanzella, E., Cristiani, S., Dickinson,
M., et al., 2005, A\&A, 434, 53
 
\bibitem[Vanzella et al.(2006)]{vanz06} Vanzella, E., Cristiani, S., Dickinson,
M., et al., 2006, A\&A, 454, 423

\bibitem[Vanzella et al.(2008)]{vanz08} Vanzella, E., Cristiani, S., Dickinson, M., 
Giavalisco, M., et al., 2008, A\&A, 478, 83

\bibitem[Vanzella et al.(2009)]{vanz09}
Vanzella, E., Giavalisco, M., Dickinson, M., Cristiani, S., Nonino, M., et al., 2009, \apj, 695, 1163

\bibitem[Verhamme et al.(2006)]{ver06} Verhamme, A., Schaerer, D., Maselli, A.,
2006, A\&A, 460, 397

\bibitem[Verhamme et al.(2008)]{ver08} Verhamme, A., Schaerer, D., Atek A., Tapken, C.,
2008, A\&A, 491, 89

\bibitem[Villar-Martin et al.(2004)]{VM04}
Villar-Martin, M., Cerviño, M., González Delgado, R. M., 2004, MNRAS, 355, 1132

\bibitem[Wiklind et al.(2008)]{wik08} Wiklind, T., Dickinson, M., Ferguson, H. C., 
Giavalisco, M., Mobasher, B., Grogin, N. A., Panagia, N., 2008, \apj, 676, 781

\bibitem[Wolfe et al. (2005)]{wolfe05}
Wolfe, A. M., Gawiser, E., Prochaska, J. X., 2005, ARA\&A, 43, 861W

\bibitem[Wuyts et al.(2008)]{wuyts08}
Wuyts, Stijn, Labb\`e, Ivo, Schreiber, Natascha M. Forster, Franx, Marijn, et al., 2008, \apj, 682, 985

\bibitem[Yan et al. (2005)]{yan05}
Yan, H., Dickinson, M., Stern, D., Eisenhardt, P. R. M., Chary, R., Giavalisco, M., et al., 2005, \apj, 634, 109

\end{thebibliography}
\end{document}